\documentclass[aps,prb,eqsecnum,twocolumn,citeautoscript,floatfix,showpacs]{revtex4}
\usepackage{pstricks}
\usepackage{amsmath,amssymb,braket}
\usepackage{graphicx,mathrsfs,textcomp,txfonts,wasysym}
\usepackage{bbold}

\newcommand{\axd}[1]{a_{#1}^{\dagger}}
\newcommand{\Ax}[1]{A_{#1}}
\newcommand{\Axd}[1]{A_{#1}^{\dagger}}
\newcommand{\bx}[1]{b_{#1}}
\newcommand{\bxd}[1]{b_{#1}^{\dagger}}
\newcommand{\Bx}[1]{B_{#1}}
\newcommand{\Bxd}[1]{B_{#1}^{\dagger}}

\newcommand{\cx}[1]{c_{#1}}
\newcommand{\cxd}[1]{c_{#1}^{\dagger}}
\newcommand{\Cx}[1]{C_{#1}}
\newcommand{\Cxd}[1]{C_{#1}^{\dagger}}

\newcommand{\bJ}{\mathbf{J}}
\newcommand{\bj}{\mathbf{j}}
\newcommand{\bp}{\mathbf{p}}

\newcommand{\one}{\mathbb{1}}
\newcommand{\Nbar}{\bar{N}}
\newcommand{\nbar}{\bar{n}}

\begin{document}
\title{Exact ground states and correlation functions of chain and 
ladder models of\\ interacting hardcore bosons or spinless fermions}
\author{Siew-Ann Cheong}
\affiliation{Laboratory of Atomic and Solid State Physics, Cornell
University, Ithaca, New York 14853-2501, USA}
\affiliation{Cornell Theory Center, Cornell University, Ithaca, New
York 14853-3801, USA}
\affiliation{Division of Physics and Applied Physics, School of
Physical and Mathematical Sciences, Nanyang Technological University, 21 Nanyang Link,
Singapore 637371, Republic of Singapore}
\author{Christopher L. Henley}
\affiliation{Laboratory of Atomic and Solid State Physics, Cornell
University, Itahca, New York 14853-2501, USA}
\date{\today}

\begin{abstract}
By removing one empty site between two occupied sites, we map the
ground states of chains of hardcore bosons and spinless fermions with
infinite nearest-neighbor repulsion to ground states of chains of
hardcore bosons and spinless fermions without nearest-neighbor
repulsion respectively, and ultimately in terms of the one-dimensional
Fermi sea.  We then introduce the intervening-particle expansion,
where we write correlation functions in such ground states as a
systematic sum over conditional expectations, each of which can be
ultimately mapped to a one-dimensional Fermi-sea expectation.  Various
ground-state correlation functions are calculated for the bosonic and
fermionic chains with infinite nearest-neighbor repulsion, as well as
for a ladder model of spinless fermions with infinite nearest-neighbor
repulsion and correlated hopping in three limiting cases.  We find
that the decay of these correlation functions are governed by
surprising power-law exponents.

\end{abstract}


\maketitle

\section{Introduction}

Exact solutions hold a special place in the theoretical condensed
matter physics of interacting electron systems.  Although they can be
obtained only for very specific models, these proved to be very useful
in understanding the behaviour of more general models of interacting
electrons, or informing us of novel physics that we would otherwise
not suspect from approximate treatments.  In particular, our present
paradigm of two universality classes, Fermi liquids versus Luttinger
liquids, for low-dimensional systems of interacting fermions came out
of exact solutions showing separation of the charge and spin degrees
of freedom \cite{tomonaga50,luttinger63,luther74,emery79}.  

In this paper, we report further surprises coming out of the exact
solution of models of hardcore bosons and spinless fermions with
infinite nearest-neighbor repulsion \cite{cheong07}.  We consider chain models
\begin{equation}\label{eqn:HBHC}
\begin{aligned}
H_{tUV}^{(c, b)} &= -t\sum_j\left[\Bxd{j}\Bx{j+1} + \Bxd{j+1}\Bx{j}\right] +
U\sum_j N_j (N_j - \one) \\
&\quad\ + \expandafter V\sum_j N_j N_{j+1}, \\
H_{tV}^{(c, f)} &= -t\sum_j\left[\Cxd{j}\Cx{j+1} + \Cxd{j+1}\Cx{j}\right] + 
V\sum_j N_j N_{j+1},
\end{aligned}
\end{equation}
of hardcore bosons ($U \to \infty$) and spinless fermions, as well as
a ladder model
\begin{equation}
\label{eqn:anisotropicextendedHubbardcorrelatedhop}
\begin{split}
H_{t_{\parallel} t_{\perp} t' V}^{(l, f)} &=
-t_{\parallel}\sum_{i=1,2}\sum_j \left(\Cxd{i,j} \Cx{i,j+1} +
\Cxd{i,j+1} \Cx{i,j}\right) \\
&\quad\ -\expandafter
t_{\perp}\sum_j \left(\Cxd{1,j} \Cx{2,j} +
\Cxd{2,j} \Cx{1,j}\right) \\
&\quad\ -\expandafter
t' \sum_j \left(\Cxd{1,j}N_{2,j+1}\Cx{1, j+2} +
\Cxd{1,j+2} N_{2,j+1} \Cx{1,j}\right) \\
&\quad\ -\expandafter
t' \sum_j \left(\Cxd{2,j} N_{1,j+1} \Cx{2, j+2} +
\Cxd{2,j+2} N_{1,j+1} \Cx{2,j}\right) \\
&\quad\ +\expandafter V\sum_i\sum_j N_{i,j} N_{i,j+1} +
V\sum_i\sum_j N_{i,j} N_{i+1,j},
\end{split}
\end{equation}
of spinless fermions.  In this ladder model\cite{zhang01}, the correlated
hopping $-t'\Cxd{i,j}N_{i'\neq i,j+1}\Cx{i,j+2}$ is the simplest term we can
introduce to blatantly favor the emergence of superconducting order.

Throughout this paper, we will specialize to the limit of infinite
onsite repulsion $U \to \infty$ and infinite nearest-neighbor
repulsion $V \to \infty$.  More precisely, we admit only
configurations in which each site can be occupied by at most one
particle, with no simultaneous occupation of nearest-neighbor sites.
We will show how the \emph{nearest-neighbor excluded chain models}
can be mapped to the \emph{nearest-neighbor included chain models}
\begin{equation}\label{eqn:HbHc}
\begin{aligned}
H_{tU}^{(c, b)} &= -t\sum_j\left[\bxd{j}\bx{j+1} + \bxd{j+1}\bx{j}\right] +
U\sum_j n_j (n_j - \one), \\
H_t^{(c, f)} &= -t\sum_j\left[\cxd{j}\cx{j+1} + \cxd{j+1}\cx{j}\right],
\end{aligned}
\end{equation}
and ultimately solve for the ground states of the former in terms of
the one-dimensional Fermi sea.  We will also show how the ladder model
can be solved exactly in three limiting cases, by mapping their ground
states to those of the chain models given in Eq.~\eqref{eqn:HBHC} and
Eq.~\eqref{eqn:HbHc}.  These analytical results were used to guide a
density-matrix analysis of correlations for the ladder model, first
using the exactly diagonalized ground states \cite{cheong09}, and
later using the density-matrix renormalization group \cite{muender09}. 

For the rest of this paper, we will consistently use uppercase letters
$\Bx{j}$ and $\Bxd{j}$ ($\Cx{j}$ and $\Cxd{j}$) to denote hardcore
boson (spinless fermion) annihilation and creation operators on
nearest-neighbor excluded chains, and lowercase letters $\bx{j}$ and
$\bxd{j}$ ($\cx{j}$ and $\cxd{j}$) to denote hardcore boson (spinless
fermion) annihilation and creation operators.  Similarly, $N_j =
\Bxd{j}\Bx{j}$ (or $N_j = \Cxd{j}\Cx{j}$) and $n_j = \bxd{j}\bx{j}$
(or $n_j = \cxd{j}\cx{j}$) are the hardcore boson (spinless fermion)
occupation number operator on the nearest-neighbor excluded and
nearest-neighbor included chains respectively.  Hereafter, we will
also use \emph{excluded} to refer to all quantities associated with
the nearest-neighbor excluded chain, and \emph{ordinary} to refer to
all quantities associated with the nearest-neighbor included chain.

Our paper will be organized as follows: in Sec.~\ref{sect:maptech}, we
will describe an analytical map that establishes a one-to-one
correspondence between the Hamiltonian matrices of the excluded and
ordinary chains of hardcore bosons and spinless fermions, before
developing a systematic expansion that would allow us to calculate
ground-state expectations in bosonic and fermionic excluded chains.
We then present and analyze in Sec.~\ref{sect:correlationsnnebosons}
correlation functions calculated using the analytical tools developed
in Sec.~\ref{sect:maptech} for excluded chains of hardcore bosons and
spinless fermions.  Following this, we write down in
Sec.~\ref{sect:laddermodels} the exact ground states of the ladder
model given in Eq.~\eqref{eqn:anisotropicextendedHubbardcorrelatedhop}
in three limiting cases, and calculate various ground-state
correlation functions, before summarizing our results and discuss the
interesting physics they imply in Sec.~\ref{sect:conclusions}.

\section{Mappings and Techniques}
\label{sect:maptech}

In Sec.~\ref{sect:nnetonni}, we establish a one-to-one correspondence
between states of the nearest-neighbor excluded and nearest-neighbor
included chains.  We explain how the Hamiltonian matrices of the two
chains, and hence their energy spectra, are identical to one another.
In the infinite-chain limit, we then show how we can write the ground
state of the excluded chain in terms of the ground state of the
ordinary chain, and ultimately be written in terms of the
one-dimensional Fermi sea.  In Sec.~\ref{sect:nnegsexp}, we show how
the ground-state expectation between two local operators can be
calculated for the excluded chain, by writing it as a systematic sum
over conditional expectations, each of which associated with a fixed
configuration of intervening particles.

For the sake of definiteness, let us consider open chains of a finite
length $L$ and total particle number $P$.  Sites on these chains are
indexed by $j = 1, \dots, L$.  Since the models in
Eq.~\eqref{eqn:HBHC} and Eq.~\eqref{eqn:HbHc} conserve $P$, the
infinite-chain limit is obtained by letting $L \to \infty$, keeping
the density of particles $\Nbar = P/L$ fixed.  Ultimately, the results
we present in this section will not depend on what boundary conditions
we impose on the chain (which is what we would expect in the infinite
chain limit).

For convenience, we establish some notations to cover boson and
fermion cases together in the same formula.  Let us call
\begin{equation}
\ket{\mathbf{J}} \equiv 
\Axd{j_1}\Axd{j_2}\cdots\Axd{j_P}\ket{0}_L,
\end{equation} 
an excluded configuration, where $A = B$ for hardcore bosons, $A = C$
for spinless fermions, and the sites $0 < j_1 < j_2 < \cdots < j_P
\leq L$ are such that $j_{p + 1} > j_p + 1$.  We will also employ the
labels $\alpha$ and $\beta$ for distinct $P$-particle configurations
$\ket{\bJ^{\alpha}}$ and $\ket{\bJ^{\beta}}$, i.e. the $P$-particle
configurations $\{j_1^{\alpha}, j_2^{\alpha}, \dots, j_P^{\alpha}\}$
and $\{j_1^{\beta}, j_2^{\beta}, \dots, j_P^{\beta}\}$ differ in at
least one site.

Similarly, let us call
\begin{equation}
\ket{\mathbf{j}} \equiv \axd{j_1}\axd{j_2}\cdots\axd{j_P}\ket{0}_L, 
\end{equation}
an ordinary configuration, where $a = b$ for hardcore bosons, $a = c$ for
spinless fermions, and the the sites $0 < j_1 < j_2 < \cdots < j_P
\leq L$ are such that $j_{p + 1} \geq j_p + 1$.  The labels $\alpha$ and
$\beta$ will again denote distinct $P$-particle configurations
$\ket{\bj^{\alpha}}$ and $\ket{\bj^{\beta}}$.  We will also
consistently denote the Hamiltonian of an excluded chain by $H_A$,
where $H_A = H_{tUV}^{(c,b)}$ for hardcore bosons, $H_A =
H_{tV}^{(c,f)}$ for spinless fermions, and the Hamiltonian of an
ordinary chain by $H_a$, where $H_a = H_{tU}^{(c,b)}$ for hardcore
bosons, $H_a = H_t^{(c,f)}$ for spinless fermions.

\subsection{Mapping Between the Excluded and Ordinary Chains}
\label{sect:nnetonni}

\begin{figure}[htbp]
\centering
\includegraphics[width=\linewidth]{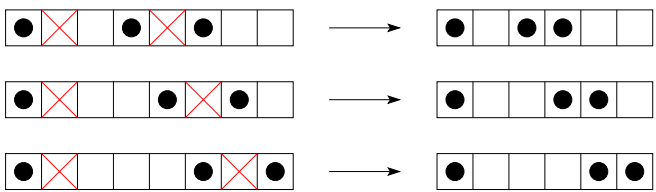}
\caption{Schematic diagram illustrating how we map from $P$-particle
configurations on an open excluded chain of length $L$ to $P$-particle
configurations on an open ordinary chain of length $L' = L - P + 1$,
by deleting one empty site to the right of a particle that is not the
rightmost particle.}
\label{fig:rightexclusionmap}
\end{figure}

In this subsection, our goal is to establish the one-to-one
correspondence between states of the excluded and ordinary chains, and
to show that as matrices, the Hamiltonians \eqref{eqn:HBHC} and
\eqref{eqn:HbHc} are identical.  To do this, let us note that an
excluded chain with $L$ sites has fewer $P$-particle states than an
ordinary chain of $L$ sites, because of the infinite nearest-neighbor
repulsion.  Therefore, we can form a one-to-one correspondence between
excluded and ordinary states only if the length $L'$ of the ordinary
chain is shorter than $L$.  There are several ways to systematically
map excluded configurations to ordinary configurations: we can (i)
delete the site to the right of every particle, if it is not the
rightmost particle; or (ii) delete the site to the left of every
particle, if it is not the leftmost particle.  We can easily check
that these maps produce the same ordinary configurations for finite
open chains.  We expect this to hold true even as we go to the
infinite chain limit.  For the rest of this paper, we will adopt
\emph{right-exclusion map}
\begin{equation}
\Axd{j_1}\Axd{j_2+1}\cdots\Axd{j_P + P - 1}\ket{0} \mapsto
\axd{j_1}\axd{j_2}\cdots\axd{j_P}\ket{0}
\end{equation}
that maps a $P$-particle configuration on an open excluded chain of
length $L$ to a $P$-particle configuration on an open ordinary chain
of length $L'$.  The empty site to the right of each occupied site in
the open excluded chain is deleted, to give a corresponding
configuration for an open ordinary chain.\footnote{A similar mapping
is possible in the case of periodic boundary conditions, as has been
discussed in Ref.~\onlinecite{cheong07}; this is pertinent to the
detailed analysis of exact diagonalizations, as applied in
Ref.~\onlinecite{cheong09}.  A significant complication of that case is that
the mapping is no longer one-to-one.  Instead, the mapping relates
Bloch states constructed as linear combinations of configurations
related by translation symmetry.  The effective length for periodic
boundary conditions is $L' = L - P$.} As illustrated in
Fig.~\ref{fig:rightexclusionmap}, we do not delete any empty site to
the right of the $P$th particle, and thus the effective length of the
open ordinary chain is $L' = L - P + 1$.  This nearest-neighbor
exclusion map was first used by Fendley to map a supersymmetric chain
of spinless fermions to the $XXZ$ chain \cite{fendley03}.  It tells us
that an excluded chain with density
\begin{equation}
\Nbar = \frac{P}{L}
\end{equation}
gets mapped to an ordinary chain with density
\begin{equation}
\nbar = \frac{P}{L'} = \frac{P}{L - P + 1} = \frac{\Nbar}{1 - \Nbar +
(1/P)}.
\end{equation}
Thus, in the limit of $L, P \to \infty$,
\begin{equation}\label{eqn:Nbarnbar}
\nbar = \frac{\Nbar}{1 - \Nbar}.
\end{equation}

For $P$-particle excluded configurations $\ket{\bJ^{\alpha}}$ and
$\ket{\bJ^{\beta}}$, the matrix element $\braket{\bJ^{\alpha} | H_A |
\bJ^{\beta}}$ is nonzero only when $\ket{\bJ^{\alpha}}$ and
$\ket{\bJ^{\beta}}$ can be obtained from one another by a single
particle hopping to the left or the right.  When this is so, the ordinary
$P$-particle configurations $\ket{\bj^{\alpha}}$ and
$\ket{\bj^{\beta}}$ they map to under the right-exclusion map are also
related to each other by a single particle hop.  Thus, we have
\begin{equation}\label{eqn:HabHAB}
\braket{\bJ^{\alpha}|H_A|\bJ^{\beta}} = -t =
\braket{\bj^{\alpha}|H_a|\bj^{\beta}}.
\end{equation}
Since there is a one-to-one correspondence between $P$-particle
configurations on an open excluded chain and $P$-particle
configurations on an open ordinary chain, Eq.~\eqref{eqn:HabHAB} tells
us that $H_A$ and $H_a$ are identical as matrices in their respective
configurational bases.  We therefore conclude that the $P$-particle
energy spectra of the two chains coincide, and that there is a
one-to-one correspondence between the energy eigenstates,
$\{\ket{\Psi}\}$ for the excluded chain, and $\{\ket{\Psi'}\}$ for the
ordinary chain.  That is, if $\ket{\bJ} \mapsto \ket{\bj}$, and
$\ket{\Psi} \mapsto \ket{\Psi'}$, then $\ket{\bJ}$ and $\ket{\bj}$
have the same amplitudes in $\ket{\Psi}$ and $\ket{\Psi'}$
respectively.  This result has profound implications on the
thermodynamics of the two chains, as well as that of the ladder model
in the three limiting cases described in
Sec.~\ref{sect:briefthreelimits}, because their partition functions
are the same.  However, for the rest of this paper, we limit ourselves
to the ground states of the infinite excluded and ordinary chains, as
well as those of the infinite ladder.

\subsection{Ground-State Expectations of the Excluded Chain}
\label{sect:nnegsexp}

In this subsection, we explain how the expectation $\braket{O}$ of an
observable $O$ in the ground state of the excluded chain can be
computed, by relating it to the expectation $\braket{O'}$ in the
ground state of the ordinary chain, for an appropriately chosen
observable $O'$ satisfying some basic correspondence requirements that
we shall outline.  Specifically, we are interested in the correlations
$\braket{O_1 O_2}$ between two local observables $O_1$ and $O_2$
separated by a distance $r$ within the excluded-chain ground state.
However, the right-exclusion map maps excluded matrix elements
$\braket{\bJ^{\alpha} | O_1 O_2 | \bJ^{\beta}}$ to ordinary matrix
elements $\braket{\bj^{\alpha} | O'_1 O'_2 | \bj^{\beta}}$ in which
$O'_1$ and $O'_2$ are separated by varying separations.  To deal with
this problem, we develop a method of intervening-particle expansion
involving a sum over conditional expectations.

To begin, let us look at the ground states
\begin{equation}
\begin{aligned}
\ket{\Psi_0} &= \sum_{\{\bJ\}} \Psi_0(\bJ) \ket{\bJ}, \\
\ket{\Psi'_0} &= \sum_{\{\bj\}} \Psi'_0(\bj) \ket{\bj}
\end{aligned}
\end{equation}
of the excluded and ordinary chains respectively.  To take advantage
of the equality of amplitudes, $\Psi_0(\bJ) = \Psi'_0(\bj)$ if
$\ket{\bJ} \mapsto \ket{\bj}$ under the right-exclusion map, we want
$\braket{\bj^{\alpha}| O' | \bj^{\beta}}$ to have a simple relation
with $\braket{\bJ^{\alpha}| O | \bJ^{\beta}}$.  While it is possible
to pick $O'$ such that $\braket{\bj^{\alpha} | O' | \bj^{\beta}} =
\braket{\bJ^{\alpha} | O |\bJ^{\beta}}$ for all $\alpha$ and $\beta$,
we find that it is more convenient to pick $O'$ such that
\begin{equation}\label{eqn:simplerrelation}
\frac{1}{\Nbar}\,\braket{\bJ^{\alpha} | O | \bJ^{\beta}} =
\frac{1}{\nbar}\,\braket{\bj^{\alpha} | O' | \bj^{\beta}}.
\end{equation}
For example, if $O = N_j = \Cxd{j}\Cx{j}$, we can pick the
corresponding observable to be $O' = n_j = \cxd{j}\cx{j}$, in which
case we find that
\begin{equation}
\begin{aligned}
\braket{\bJ^{\alpha} | N_j | \bJ^{\beta}} &= \Nbar\,
\delta_{\alpha\beta}, \\
\braket{\bj^{\alpha} | n_j | \bj^{\beta}} &= \nbar\,
\delta_{\alpha\beta},
\end{aligned}
\end{equation}
which satisfies Eq.~\eqref{eqn:simplerrelation} trivially.  We call
$O$ and $O'$ a \emph{corresponding pair of observables}, if
Eq.~\eqref{eqn:simplerrelation} is satisfied for all $\alpha$ and
$\beta$, allowing us to write the very simple relation
\begin{equation}\label{eqn:ON1On1}
\frac{1}{\Nbar}\braket{O} = \frac{1}{\nbar}\braket{O'}
\end{equation}
between their ground-state expectations.

Since we are mostly interested in correlation functions within the
excluded chain ground state, let us look at expectations of the
product form $\braket{O_j O_{j+r}}$, where $O_j$ acts locally about
site $j$, and $O_{j+r}$ acts locally about site $j + r$.  Becasue the
number of particles $p$ between sites $j$ and $j+r$ varies from
excluded configuration to excluded configuration, these sites get
mapped by the right-exclusion map to sites on the ordinary chain with
varying separations $r - p$.  Therefore, to calculate the excluded
chain ground-state expectation $\braket{O_j O_{j+r}}$ in terms of
ordinary chain ground-state expectations, we first write down an
\emph{intervening-particle expansion}
\begin{equation}\label{eqn:interpartexp}
\braket{O_j O_{j+r}} = \sum_{\bp} \braket{O_j Q_{\bp} O_{j+r}},
\end{equation}
where $\braket{O_j Q_{\bp} O_{j+r}}$ are conditional expectations.
Here $\bp$ is a vector of occupation numbers within the intervening
sites, and $Q_{\bp}$ is a string of factors, each of which is either
$N_{j + s}$ or $(\one - N_{j + s})$, $1 \leq s \leq r - 1$.  The sum
is over all possible ways to have intervening particles between $O_j$
and $O_{j+r}$.  For each excluded term $\braket{O_j Q_{\bp} O_{j+r}}$
in Eq.~\eqref{eqn:interpartexp}, we then write down the corresponding
ordinary expectation $\braket{O'_j Q'_{\bp'} O'_{j+r-p}}$, and
thereafter sum over all corresponding ordinary expectations,
\begin{equation}
\braket{O_j O_{j+r}} = \frac{\Nbar}{\nbar}
\sum_{\bp'} \braket{O'_j Q'_{\bp'} O'_{j+r-p}},
\end{equation}
making use of Eq.~\eqref{eqn:ON1On1}.  The vector $\bp'$ of occupation
numbers is obtained from $\bp$ using the right-exclusion map, and
contains the same number $p$ of occupied intervening sites.

To illustrate how the corresponding expectations $\braket{O'_j
Q'_{\bp'} O'_{j+r-p}}$ can be constructed, let us write
Eq.~\eqref{eqn:interpartexp} out explicitly as
\begin{equation}\label{eqn:extendedhardcoreexpansion}
\begin{split}
\braket{O_{j} O_{j+r}} &=
\braket{O_j (\one - N_{j+1})\cdots(\one - N_{j+r-1}) O_{j+r}} + {} \\
&\quad\ \braket{O_j N_{j+1}\cdots(\one - N_{j+r-1}) O_{j+r}} + 
\cdots + {} \\
&\quad\ \braket{O_j (\one - N_{j+1})\cdots N_{j+r-1} O_{j+r}} + {} \\
&\quad\ \braket{O_j N_{j+1}N_{j+2}\cdots
(\one - N_{j+r-1}) O_{j+r}} + \cdots + {} \\
&\quad\ \braket{O_j (\one - N_{j+1})\cdots 
N_{j+r-2} N_{j+r-1} O_{j+r}} + \cdots + {} \\
&\quad\ \braket{O_j N_{j+1}N_{j+2}\cdots N_{j+r-1} O_{j+r}},
\end{split}
\end{equation}
each of which contains intervening particles at fixed sites.  We call
terms in the expansion with $p$ intervening $N_j$'s the
\emph{$p$-intervening-particle expectations}.  Because of
nearest-neighbor exclusion, most of the terms in
Eq.~\eqref{eqn:extendedhardcoreexpansion} vanish.

Next, we map each conditional excluded expectation in
Eq.~\eqref{eqn:extendedhardcoreexpansion} to a corresponding
conditional ordinary expectation following the simple rules given below:
\begin{enumerate}

\item \textbf{Nearest-neighbor exclusion.}  To ensure that we do not
violate nearest-neighbor exclusion, we make the assignment
\begin{equation}\label{eqn:BBexclude}
\Axd{j+s}\Axd{j+s+1} = 0 = \Ax{j+s}\Ax{j+s+1}.
\end{equation}
Note that this is intended not as a statement on the operator algebra, 
but as a mere bookkeeping device for evaluating expectations.
The assignment
\begin{equation}\label{eqn:BNexclude}
\Axd{j+s} N_{j+s+1} = 0 = N_{j+s} \Ax{j+s+1}
\end{equation}
follows from Eq.~\eqref{eqn:BBexclude}.

\item \textbf{Right-exclusion map.}  The right-exclusion map described
in Section \ref{sect:nnetonni} is then implemented by making the
substitution
\begin{equation}\label{eqn:Brightmerge}
\Axd{j+s}(\one - N_{j+s+1}) \mapsto \axd{j+s}.
\end{equation}
The assignment
\begin{equation}\label{eqn:interveningpairmap}
N_{j+s} (\one - N_{j+s+1}) = n_{j+s}
\end{equation}
follows from Eq.~\eqref{eqn:Brightmerge}.

\item \textbf{Re-indexing.}  Because the right-exclusion map in
Eq.~\eqref{eqn:Brightmerge} merges the occupied site $j+s$ and the
empty site $j+s+1$ to its right, operators to the right of site
$j+s+1$ must be re-indexed.  The index $j+s$ on the excluded chain
becomes
\begin{equation}
j + s - \sum_{s'=0}^{s - 1} N_{j + s'}
\end{equation}
on the ordinary chain.  Thus, two ending operators $r$ sites apart in
the $p$-intervening-particle excluded expectation becomes $r - p$
sites apart in the corresponding $p$-intervening-particle ordinary
expectation.

\end{enumerate}

\section{Correlations in the Bosonic and Fermionic Excluded Chains}
\label{sect:correlationsnnebosons}

In this section, we make use of the tools developed in
Sec.~\ref{sect:maptech} to calculate three simple correlation
functions within the ground states of the excluded chains of hardcore
bosons and spinless fermions.  In general, the intervening-particle
expansion for excluded chain ground-state correlations must be
evaluated numerically (even when each ordinary chain ground-state
expectations in the sum can be expressed in closed form), keeping in
mind a excluded chain with density $\Nbar$ maps to an ordinary chain
with density $\nbar = \Nbar/(1 - \Nbar)$ (see
Eq.~\eqref{eqn:Nbarnbar}).

In Sec.~\ref{sect:FLchain}, Sec.~\ref{sect:CDWchain}, and
Sec.~\ref{sect:SCchain}, we show numerical results for the two-point
functions $\braket{\Bxd{j}\Bx{j+r}}$ and $\braket{\Cxd{j}\Cx{j+r}}$,
and the four-point functions $\braket{N_j N_{j+r}}$,
$\braket{\Bxd{j}\Bxd{j'}\Bx{j+r}\Bx{j'+r}}$ and
$\braket{\Cxd{j}\Cxd{j'}\Cx{j+r}\Cx{j'+r}}$ respectively.  For the
sake of easy reference, we will call the two-point functions
$\braket{\Bxd{j}\Bx{j+r}}$ and $\braket{\Cxd{j}\Cx{j+r}}$ Fermi-liquid
(FL) type correlation functions, even though their spatial structures
depend on particle statistics.  We will also call the four-point
functions $\braket{N_j N_{j+r}} = \braket{\Bxd{j}\Bx{j}
\Bxd{j+r}\Bx{j+r}}, \braket{\Cxd{j}\Cx{j} \Cxd{j+r}\Cx{j+r}}$ CDW type
correlation functions, and the the four-point functions
$\braket{\Bxd{j}\Bxd{j'}\Bx{j+r}\Bx{j'+r}}$ and
$\braket{\Cxd{j}\Cxd{j'}\Cx{j+r}\Cx{j'+r}}$ SC type correlation
functions.  Both CDW and SC type correlations are identical for
hardcore bosons and spinless fermions on the excluded chain, but the
latter has the `superconducting' interpretation only for fermions.

In Sec.~\ref{sect:FLchain}, we will also explain in detail how
nonlinear curve fits of the numerical correlations to reasonable
asymptotic forms as a function of $r$ are done.  Based on results from
the nonlinear curve fits, we show how the Luttinger's theorem
manifests itself, and how meaningful power-law exponents can be
extracted.  Similar analyses are done in Sec.~\ref{sect:CDWchain} and
Sec.~\ref{sect:SCchain}, as well as in Sec.~\ref{sect:laddermodels}
for the three limiting ground states of the ladder model.

\subsection{FL Correlations}
\label{sect:FLchain}

In the intervening-particle expansions of the two-point functions
$\braket{\Bxd{j}\Bx{j+r}}$ and $\braket{\Cxd{j}\Cx{j+r}}$, the
nonvanishing terms map to $p$-intervening-particle expectations of the
form $\braket{\bxd{j} \prod_{j_p} n_{j_p} \prod_{j_h} (\one - n_{j_h})
\bx{j + r - p}}$ and $\braket{\cxd{j} \prod_{j_p} n_{j_p} \prod_{j_h}
(\one - n_{j_h}) \cx{j + r - p}}$.  Both can be evaluated in terms of
two-point functions
\begin{equation}
\braket{\cxd{i}\cx{j}} = \frac{\sin\nbar\pi|i - j|}{\pi|i - j|}
\end{equation}
of the one-dimensional Fermi sea, after invoking the Jordan-Wigner
transformation (see Appendix \ref{sect:JWtrans}) for the former.

\begin{figure}[htbp]
\centering
\includegraphics[width=\linewidth,clip=true]{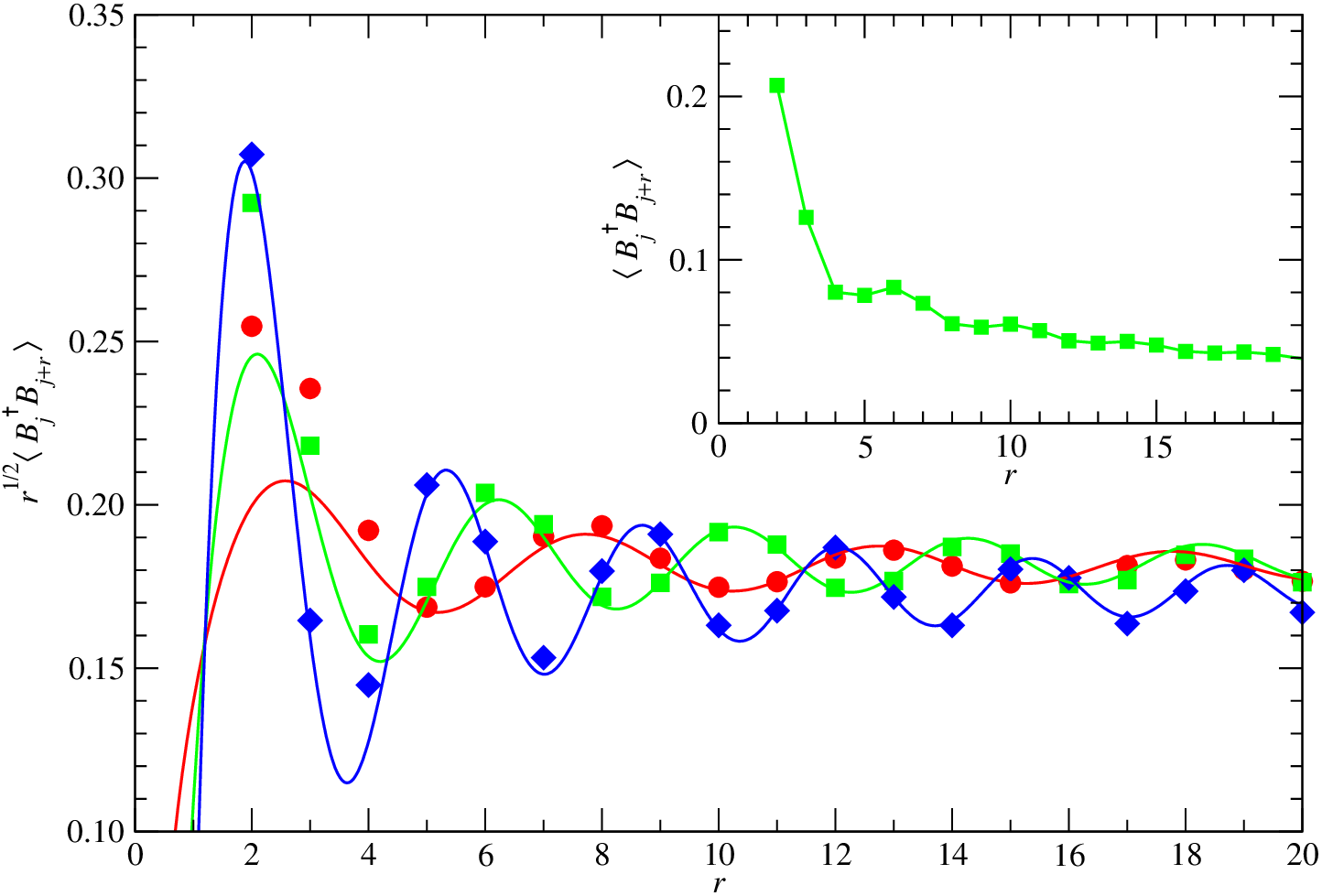}
\caption{(Color online)  Plot of $\sqrt{r}\braket{\Bxd{j}\Bx{j+r}}$ as
a function of $r$ for the particle densities $\Nbar = 0.20$ (red
circles), $\Nbar = 0.25$ (green squares), $\Nbar = 0.30$ (blue
diamonds).  The colored curves shown are nonlinear curve fits of
$\sqrt{r} \braket{\Bxd{j} \Bx{j+r}}$ to the asymptotic form $A_0 +
A_1\, r^{-\alpha'_1}\cos(2 k_F r + \phi_1)$.  (Inset) Plot of
$\braket{\Bxd{j}\Bx{j+r}}$ as a function of $r$, showing that it
consists of a simple power law part and an oscillatory power law
part.}
\label{figure:flbN20N25N30}
\end{figure}

As shown in the inset of Fig.~\ref{figure:flbN20N25N30}, the FL
correlation $\braket{\Bxd{j}\Bx{j+r}}$ was found to consists of a
simple power law part, decaying with a smaller exponent $\alpha_0$,
and an oscillatory power law part, decaying with a larger exponent
$\alpha_1$.  Multiplying $\braket{\Bxd{j}\Bx{j+r}}$ by various simple
powers of $r$, we find that $\sqrt{r}\braket{\Bxd{j}\Bx{j+r}}$
asymptotes to a constant with large $r$ (as shown in the main plot of
Fig.~\ref{figure:flbN20N25N30}), which suggests that $\alpha_0 =
\frac{1}{2}$.  This is the correlation exponent predicted by Efetov
and Larkin, in their study of the ordinary chain of hardcore bosons
\cite{efetov76}.

\begin{figure}[htbp]
\centering
\includegraphics[width=\linewidth,clip=true]{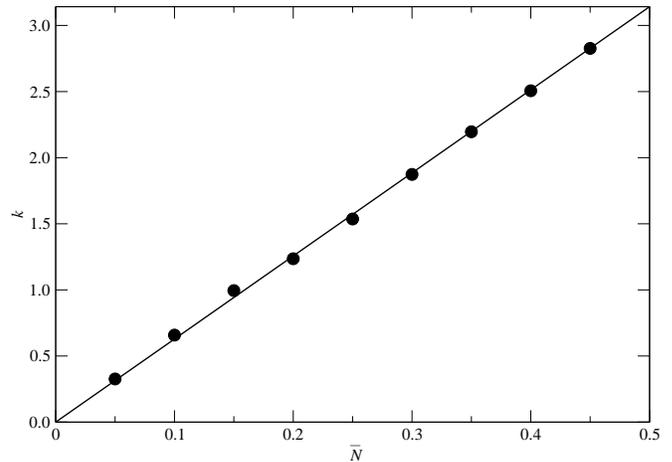}
\caption{Plot of the fitted wave number $k$ (solid circles) against
the density $\Nbar$ of the excluded chain of hardcore bosons.  The
parameter $k$ is obtained from the unrestricted nonlinear curve fit of
$\sqrt{r}\braket{\Bxd{j}\Bx{j+r}}$ to the asymptotic form $A_0 + A_1\,
r^{-\alpha'_1}\cos(k r + \phi_1)$.  The straight line is $k_F =
2\pi\Nbar$ for the Fermi wave number.}
\label{figure:flbkfit}
\end{figure}

Another important result comes from the unrestricted nonlinear curve
fits of $\sqrt{r}\braket{\Bxd{j}\Bx{j+r}}$ to the asymptotic form $A_0
+ A_1\, r^{-\alpha'_1}\, \cos(k r + \phi_1)$.  In
Fig.~\ref{figure:flbkfit} we show a plot of the fitted wave number $k$
as a function of the density $\Nbar$ of the excluded chain of hardcore
bosons.  As we can see, the fitted wave numbers fall neatly onto the
straight line $k = 2 k_F = 2\pi\Nbar$, where $k_F = \pi\Nbar$ is the
Fermi wave number.  The fact that $k_F$ appears naturally in the
numerical correlations is expected from Luttinger's theorem, which
states that the volume of the reciprocal space bounded by the
noninteracting Fermi surface is invariant quantity not affected by
interactions, and applies in both Fermi and non-Fermi liquids
\cite{luttinger60, luttinger63, oshikawa00, dzyaloshinskii03,
korshunov03, paramekanti04, liu05}.  From this point onwards, we
restrict the wave number of the oscillatory part of the correlation
functions to $k_F$, $2 k_F$, or $4k_F$ in the nonlinear curve fits.

\begin{figure}[htbp]
\centering
\includegraphics[width=\linewidth,clip=true]{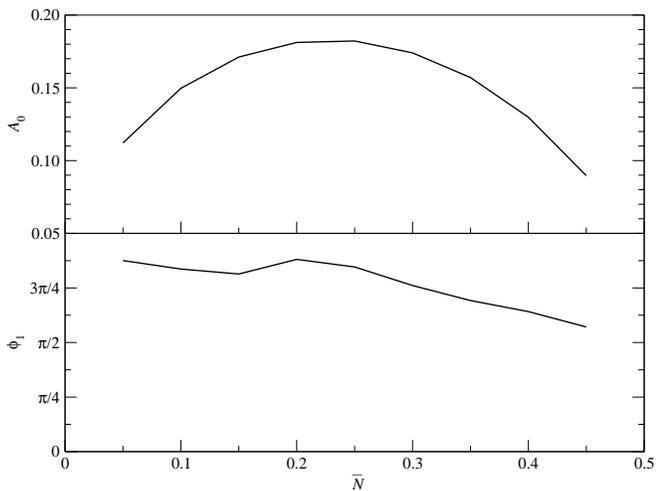}
\caption{Plot of the fitted amplitude $A_0$ (top) of the simple power
law part, and the fitted phase shift $\phi_1$ (bottom) of the
oscillatory power law part, as functions of the density $\Nbar$ of the
excluded chain of hardcore bosons.}
\label{figure:flbAphifit}
\end{figure}

From the restricted nonlinear curve fits, we find that $A_1$ is large
when $\alpha'_1$ is large, and small when $\alpha'_1$ is small.  This
suggests that neither of these parameters can be accurately determined
from our nonlinear curve fits, unless we further constrain what values
$\alpha'_1$ can take.  We also find that the quality of the nonlinear
curve fit is good when $\Nbar$ is far from $\Nbar = \frac{1}{4}$, but
deteriorates as we approach quarter filling.  This suggests important
physics in the FL correlation $\braket{\Bxd{j}\Bx{j+r}}$ near quarter
filling, which cannot be adequately accounted for by an asymptotic
form $A_0 + A_1\, r^{-\alpha'_1}\, \cos(k r + \phi_1)$.  This loss of
fit also affects the phase shift $\phi_1$, presumably to a smaller
extent, and the amplitude $A_0$ of the simple power law, to an even
smaller extent.  These two parameters are plotted as functions of the
density in Fig.~\ref{figure:flbAphifit}.  In the limit $\Nbar \to 0$,
we have essentially a dilute gas of hardcore (otherwise
noninteracting) bosons, and thus $\braket{\Bxd{j}\Bx{j+r}}$ should
include an overall factor of $\Nbar$.  Thus we expect $A_0 \to 0$ as
$\Nbar \to 0$.  In the limit $\Nbar \to \frac{1}{2}$, the excluded
chain of hardcore bosons becomes increasingly jammed, and the relevant
degrees of freedom are holes.  For a dilute chain of holes, we expect
$\braket{\Bxd{j}\Bx{j+r}}$ to be proportional to the hole density, and
thus $A_0 \to 0$ as $\Nbar \to \frac{1}{2}$.  From our numerical
results alone, it is hard to tell whether $A_0$ reaches a maximum at
$\Nbar = \frac{1}{5}$ (corresponding to a quarter-filled, $\nbar =
\frac{1}{4}$, ordinary chain of hardcore bosons) or $\Nbar =
\frac{1}{4}$ (quarter-filled excluded chain of hardcore bosons).  It
is also hard to say anything definite about the phase shift $\phi_1$,
which might in fact be constant.

\begin{figure}[htbp]
\centering
\includegraphics[width=\linewidth,clip=true]{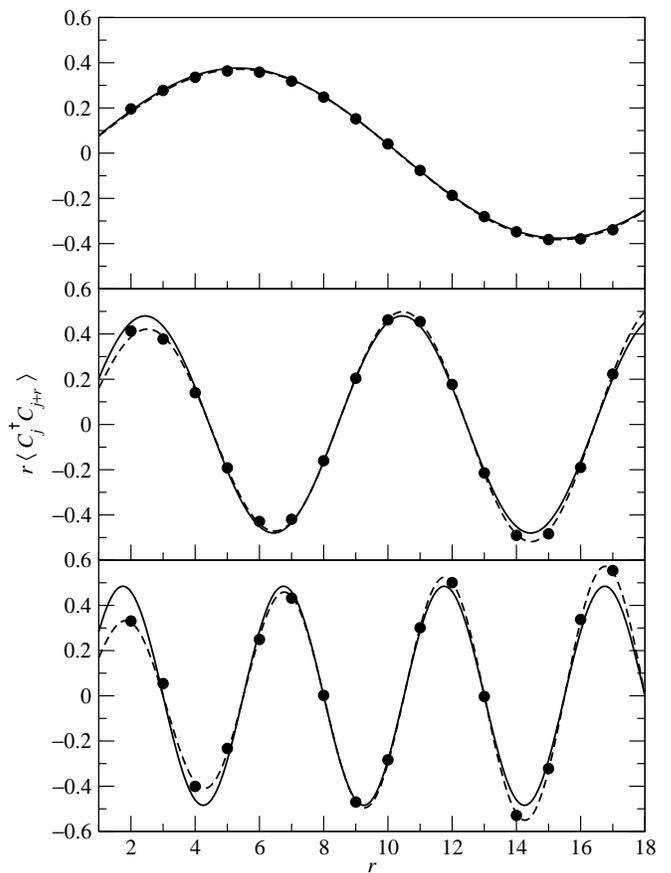}
\caption{Plots of $r\braket{ \Cxd{j} \Cx{j+r}}$ as functions of $r$
(solid circles) for particle densities $\Nbar = 0.10$ (top), $\Nbar =
0.25$ (middle), and $\Nbar = 0.40$ (bottom).  Nonlinear curve fits of
$r\braket{\Cxd{j}\Cx{j+r}}$ to the restricted asymptotic form
$A_1\cos(\pi\Nbar r + \phi_1)$ (solid curves) show systematic
deviations, which can be accounted for by a unrestricted asymptotic
form $A_1\, r^{(1 - \alpha_1)}\cos (\pi\Nbar r + \phi_1)$.}
\label{figure:flfN10N25N40}
\end{figure}

\begin{figure}[htbp]
\centering
\includegraphics[width=\linewidth,clip=true]{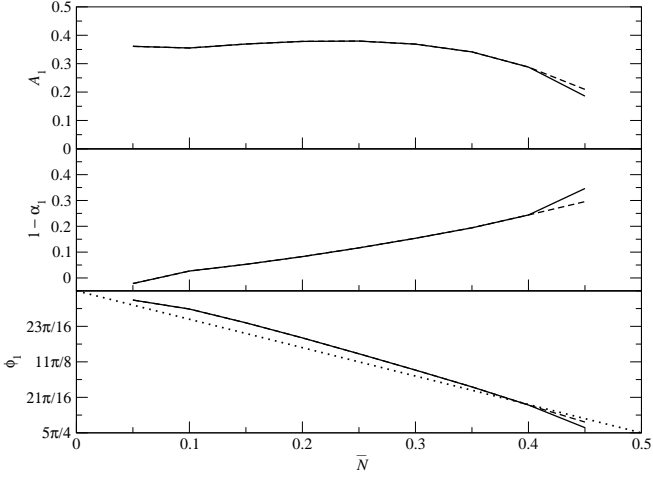}
\caption{Plot of the fitted amplitude $A_1$ (top), the fitted exponent
$\alpha_1$ (middle), and the fitted phase shift $\phi_1$ of the
leading oscillatory power-law decay in the FL correlation
$\braket{\Cxd{j}\Cx{j+r}}$, as functions of the density $\Nbar$ of the
excluded chain of spinless fermions.  In these plots,
the solid curves are for fits to $A_1\, r^{1 - \alpha_1}\,
\cos(\pi\Nbar r + \phi_1)$, whereas the dashed curves are for fits to
$A_1\, r^{1 - \alpha_1}\, \cos(\pi\Nbar r + \phi_1) + A_2\, r^{1 -
\alpha_2}\, \cos[\pi(1 - \Nbar) r + \phi_2]$.  The dotted line in the
bottom plot is a straight line from $5\pi/4$ at $\Nbar = 0$ to
$3\pi/2$ at $\Nbar = \frac{1}{2}$ to guide the eye.}
\label{figure:flffits}
\end{figure}

In contrast, the FL correlation $\braket{\Cxd{j}\Cx{j+r}}$ contains no
simple power law part.  A preliminary unrestricted nonlinear curve fit
of this correlation to the asymptotic form $A_1\, r^{-\alpha_1}\,
\cos(\pi\Nbar r + \phi_1)$ suggests that $\alpha_1 \approx 1$ for all
densities, i.e. may be universal $\alpha_1 = 1$ just as for
noninteracting spinless fermions.  However, a more careful restricted
nonlinear curve fit $r\braket{\Cxd{j}\Cx{j+r}} = A_1 \cos(\pi\Nbar r +
\phi_1)$ show systematic deviations, as shown in
Fig.~\ref{figure:flfN10N25N40}, and therefore we perform an
unrestricted fit to $r\braket{\Cxd{j}\Cx{j+r}} = A_1\, r^{1 -
\alpha_1}\, \cos(\pi\Nbar r + \phi_1)$.  The fitted parameters $A_1$,
$1 - \alpha_1$, and $\phi_1$ are shown in Fig.~\ref{figure:flffits}.

At very low densities $\Nbar \to 0$, our dilute chain of
nearest-neighbor excluding spinless fermions will behave like
noninteracting fermions, so we expect
\begin{equation}\label{eqn:lowdensitylimit}
\braket{\Cxd{j}\Cx{j+r}} \approx \frac{\sin\pi\Nbar r}{\pi r}.
\end{equation}
From our curve fits, we see that in this limit, $A_1 \to 1/\pi =
0.31831\dots$, $1 - \alpha_1 \to 0$, and $\phi_1 \to 3\pi/2$, and thus
the FL correlation does indeed go to the low density limit in
Eq.~\eqref{eqn:lowdensitylimit}.

Also, in the half-filling limit $\Nbar \to \frac{1}{2}$, the chain
become more and more congested, making it increasingly difficult to
annihilate a spinless fermion at site $j+r$, find an empty site $j$ to
create a spinless fermion, without running afoul of the
nearest-neighbor exclusion constraint.  This tells us that $A_1$ must
vanish as $\Nbar \to \frac{1}{2}$, which is hinted at in
Fig.~\ref{figure:flffits}.  However, the vanishing amplitude is only
half of the story in this limit, the other half being the rate at
which the correlation decay with increasing separation.  In fact, very
close to $\Nbar = \frac{1}{2}$, we expect the ground-state physics of
the chain of rung-fermions with infinite nearest-neighbor repulsion to
be describable in terms of a low density of holes.  Naively, we would
expect from such a low-density-of-holes argument that $\braket{\Cxd{j}
\Cx{j+r}}$ decay as $r^{-1}$.  Instead, the nonlinear curve fits at
$\Nbar_1 \lesssim \frac{1}{2}$ tells us that $\alpha_1 < 1$.

Thinking about this nearly-half-filled limit more carefully, we
realized that what we called `holes' are really domain walls
separating a region in which the spinless fermions sit on odd sites,
from a region in which the spinless fermions sit on even sites.  The
FL correlation $\braket{\Cxd{j}\Cx{j+r}}$, which can be written as a
hole-hole correlation function, then depends on how many holes there
are between $j$ and $j+r$.  The idea is that, in order to annihilate a
hole (create a spinless fermion) at site $j+r$ and create a hole
(annihilate a spinless fermion) at site $j$, we must first find a
configuration with a hole at $j+r$.  Such a configuration will have
spinless fermions at sites $j+r-2$, $j+r-4$, \dots, until we encounter
another hole at $j + r - 2s$, and then the sequence of spinless
fermions will thereafter be at sites $j + r - 2s - 1$, $j + r - 2s -
3$, \dots.  If $r$ is even, $\braket{\Cxd{j}\Cx{j+r}}$ receives nonzero
contributions only from those configurations with an even number of
intervening holes, whereas if $r$ is odd, $\braket{\Cxd{j}\Cx{j+r}}$
receives nonzero contributions only from those configurations with an
odd number of intervening holes.  This is very similar in flavor to
the intervening-particle expansion of the two-point function
$\braket{\bxd{j}\bx{j+r}}$ of a chain of ordinary hardcore bosons, except
that $\braket{\bxd{j}\bx{j+r}}$ receives positive contributions from
configurations with an even number of intervening particles, and
negative contributions from an odd number of intervening particles.
Therefore, in the limit $\Nbar \to \frac{1}{2}$, we find that the FL
correlation $\braket{\Cxd{j}\Cx{j+r}}$ maps to a string correlation of
holes.  Bosonization calculations then show that this string
correlation of holes decay as a power law, with correlation exponent
$\alpha_1 = \frac{1}{4}$ \cite{henley05}.

\subsection{CDW Correlations}
\label{sect:CDWchain}

\begin{figure}[htbp]
\centering
\includegraphics[width=\linewidth,clip=true]{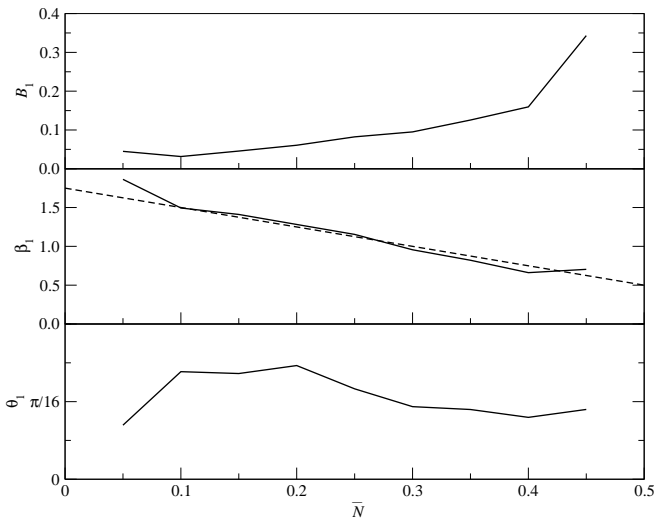}
\caption{Plot of the fitted amplitude $B_1$ (top), the fitted exponent
$\beta_1$ (middle) and the fitted phase shift $\theta_1$ (bottom) of
the leading oscillatory power law decay of the subtracted CDW
correlation $\braket{N_j N_{j+r}} - \braket{N_j}\braket{N_{j+r}}$, as
functions of the density $\Nbar$ of the excluded chain of hardcore
bosons/spinless fermions.}
\label{figure:cdwbfits}
\end{figure}

Important physics can also be learnt from the nonlinear curve fitting
of the CDW correlations $\braket{ N_j N_{j+r} } \equiv \braket{
\Bxd{j} \Bx{j} \Bxd{j+r} \Bx{j+r} } = \braket{ \Cxd{j} \Cx{j}
\Cxd{j+r} \Cx{j} }$.  First we tried to fit the subtracted CDW
correlation to the asymptotic form $\braket{N_j N_{j+r}} -
\braket{N_j} \braket{N_{j+r}} = B_1\,r^{-\beta_1}\,\cos(2\pi\Nbar r +
\theta_1)$, but found the quality of fit deteriorates as $\Nbar \to
0$, as shown in Fig.~\ref{figure:cdwbN10N25N45}.  We understand this
as follows: for $\Nbar \to 0$, the dimensionless quantity $\xi = \Nbar
r$ is small, and the poor fit indicates that $\braket{N_j N_{j+r}}$
contains contributions from a term that decays more rapidly than
$B_1\,r^{-\beta_1}\,\cos(2\pi\Nbar r + \theta_1)$.  If we assume that
this faster decaying term is a simple power law of the form $B_2\,
r^{-\beta_2}$, and fit the CDW correlation to $\braket{N_j N_{j+r}} =
B_1\,r^{-\beta_1}\,\cos(2\pi\Nbar r + \theta_1) + B_2\, r^{-\beta_2}$,
we found the quality of the nonlinear curve fit is improved, after
dropping data points $r = 2, 3$ from the fit, as shown in
Fig.~\ref{figure:cdwbN10N25N45}.  We can include the simple power law
decay term in the nonlinear curve fit throughout the entire range of
$\Nbar$, but the parameters $B_2$ and $\beta_2$ cannot be reliably
determined beyond $\Nbar = \frac{1}{4}$.  Therefore, the parameters
$B_1$, $\beta_1$, and $\theta_1$ are the only parameters that can be
reliably determined across the whole range of densities.  From
Fig.~\ref{figure:cdwbfits}, we find that $\theta_1$ changes very
little over the whole range of densities, and remains close to
$\pi/16$.  On the other hand, the leading correlation exponent
$\beta_1$ appears to be density-dependent, and is very close to being
\begin{equation}\label{eqn:beta1}
\textstyle\beta_1 = \frac{1}{2} + \frac{5}{2}\left(\frac{1}{2} - \Nbar\right).
\end{equation}

\begin{figure}[htbp]
\centering
\includegraphics[width=\linewidth,clip=true]{cdwbN10N25N45}
\caption{Nonlinear curve fits of the subtracted CDW correlation
$\braket{N_j N_{j+r}}$ (solid circles) in the bosonic/fermionic ground
states of the excluded chain, at densities $\Nbar = 0.10$ (top), $\Nbar =
0.25$ (middle), and $\Nbar = 0.45$ (bottom).  Above quarter filling
(middle and bottoms plots), the subtracted CDW correlations can be
fitted very well to the simple asymptotic form $B_1\, r^{-\beta_1}
\cos(2\pi\Nbar r + \theta_1)$ (solid curves), whereas at low densities
(top plot), the subtracted CDW correlations deviate significantly from
this simple asymptotic form.  The nonlinear curve fit improves only
after we add a simple power-law correction term $B_2\, r^{-\beta_2}$,
giving the dashed curves.}
\label{figure:cdwbN10N25N45}
\end{figure}

Just as for the FL correlation, we need to again think hard about the
hole physics of the ground-state CDW correlation very close to
half-filling.  In this limit, the CDW correlation $\braket{N_j
N_{j+r}}$ can be written in terms of the hole-density-hole-density
correlation $\braket{P_j P_{j+r}}$, where $P_j = \one - N_j$ is the
hole occupation number operator on site $j$.  Using an argument
similar to the one used for the FL correlation $\braket{\Cxd{j}
\Cx{j+r}}$ near half-filling, we realized that if $r$ is even,
$\braket{P_jP_{j+r}}$ receives nonzero contribution from
configurations with an even number of intervening holes, whereas if
$r$ is odd, $\braket{P_jP_{j+r}}$ receives nonzero contribution from
configurations with an odd number of intervening holes.  This means
that our hole-density-hole-density correlation $\braket{P_jP_{j+r}}$
must be mapped to a string correlation of a chain of noninteracting
spinless fermions.  Bosonization calculations tell us that this string
correlation decays as a power law with correlation exponent $\alpha =
\frac{1}{2}$, consistent with the conjectured behaviour,
Eq.~\eqref{eqn:beta1}, at the special value $\Nbar = \frac{1}{2}$
\cite{henley05}.

\subsection{SC Correlations}
\label{sect:SCchain}

\begin{figure}[htbp]
\centering
\includegraphics[width=\linewidth,clip=true]{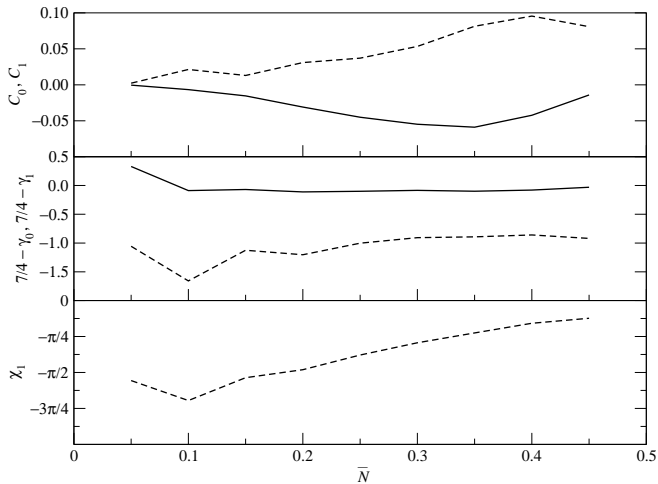}
\caption{Plot of the fitted amplitudes $C_0$ and $C_1$ (top), the
fitted exponents $\gamma_0$ and $\gamma_1$ (middle), and the fitted
phase shift $\chi_1$ of the leading simple power-law decay and the
subleading oscillatory power-law decay in the SC correlation
$\braket{\Axd{j-2}\Axd{j}\Ax{j+r}\Ax{j+r+2}}$, as functions of the
density $\Nbar$ of the excluded chain of hardcore
bosons/spinless fermions.}
\label{figure:scffits}
\end{figure}

In contrast to the FL and CDW correlations, the SC correlation
$\braket{\Axd{j-2}\Axd{j}\Ax{j+r}\Ax{j+r+2}}$ on the excluded chain
has a rather more complex structure.  The SC correlation is always
negative, and oscillations are highly suppressed, suggesting that it
is the sum of a simple power law and an oscillatory power law.  To
improve the reliability of the nonlinear curve fits, we prescale the
SC correlation by multiplying it by $r^{7/4}$.  This strange exponent
is chosen because it is closest to the rate at which the simple power
law decay for various densities.  After dropping data points for $r =
2, 3, 4$, good fits to the asymptotic form $r^{7/4} \braket{ \Axd{j-2}
\Axd{j} \Ax{j+r} \Ax{j+r+2}} = C_0\, r^{7/4 - \gamma_0} + C_1\, r^{7/4
- \gamma_1}\, \cos(2\pi\Nbar r + \chi_1)$ were obtained.  The
nonlinear curve fits were improved marginally by adding a correction
term of the form $C_2\, r^{7/4 - \gamma_2}\, \cos(4\pi\Nbar r +
\chi_2)$ (see Fig.~\ref{figure:scfN10N25N45}).  The fitted parameters
are shown in Fig.~\ref{figure:scffits} as functions of the excluded
chain density $\Nbar$.

\begin{figure}[htbp]
\centering
\includegraphics[width=\linewidth,clip=true]{scfN10N25N45}
\caption{Nonlinear curve fits of the SC correlation $r^{7/4}
\braket{\Axd{j-2} \Axd{j} \Ax{j+r} \Ax{j+r+2}}$ (solid circles) in the
bosonic/fermionic ground states of the excluded chain, at densities $\Nbar
= 0.10$ (top), $\Nbar = 0.25$ (middle), and $\Nbar = 0.45$ (bottom).
After dropping data points from $r = 2, 3, 4$, the SC correlations can
be fitted well to the simple asymptotic form $C_0\, r^{7/4 - \gamma_0}
+ C_1\, r^{7/4 - \gamma_1}\, \cos(2\pi\Nbar r + \chi_1)$ (solid
curves).  The nonlinear curve fits are marginally improved by adding a
correction term of the form $C_2\, r^{7/4 - \gamma_2}\, \cos(4\pi\Nbar
r + \chi_2)$ (dashed curves).}
\label{figure:scfN10N25N45}
\end{figure}

Unlike for the FL and CDW correlations, there are no analytical
bosonization calculations to help suggest values for the SC
correlation exponents, so we used an ad-hoc process where we imposed
trial values of the exponents, and let the nonlinear curve fitting
program find the appropriate amplitudes and phase shifts.  We found
visually that the best fit of the numerical SC correlations appears to
the mixed asymptotic form
\begin{multline}\label{eqn:SCmixed}
r^{7/4} \braket{ \Axd{j-2} \Axd{j} \Ax{j+r} \Ax{j+r+2}} = 
C'_0\, r^{-\frac{1}{8}} + {} \\
C'_1\, r^{-\frac{1}{4}}\, \cos(2\pi\Nbar r + \chi'_1) + {} \\
C'_2\, r^{-\frac{3}{2}}\, \cos(2\pi\Nbar r + \chi'_2) + 
C'_3\, r^{-\frac{7}{2}}.
\end{multline}

\section{Ladder Model}
\label{sect:laddermodels}

In this section we show how the analytical machinery developed in
Sec.~\ref{sect:maptech} can be adapted to calculate ground-state
correlations in the ladder model of interacting spinless fermions
given in Eq.~\eqref{eqn:anisotropicextendedHubbardcorrelatedhop}, in
three limiting cases where the ground states can be deduced from
simple energetic arguments.  An overview is given in
Sec.~\ref{sect:briefthreelimits}, before we move on to detailed
analyses and discussions of the three limiting cases in
Sec.~\ref{sect:infinitelystrongcorrelatedhops},
Sec.~\ref{sect:independentlegs}, and Sec.~\ref{sect:independentrungs}.
As with the chain models, we assume that the ladder is finite, with $j
= 1, \dots, L$, and subject each of its legs $i = 1, 2$ to open
boundary conditions.  Exact solution for the infinite ladder is then
obtained by taking $L \to \infty$ keeping the particle density
$\Nbar_2$ fixed.  Just as for the chain models, we expect in this
limit that the ladder exact solutions would not depend on which
boundary conditions we used.

\subsection{The Three Limiting Cases: An Overview}
\label{sect:briefthreelimits}

For the ladder model described by
Eq.~\eqref{eqn:anisotropicextendedHubbardcorrelatedhop}, with $V \to
\infty$, the ground state is determined by the two independent model
parameters, $t_{\perp}/t_{\parallel}$ and $t'/t_{\parallel}$, and the
density $\Nbar_2$.  For fixed $\Nbar_2$, the two-dimensional region in
the ground-state phase diagram is bounded by three limiting cases,
\begin{enumerate}
\renewcommand{\labelenumi}{(\roman{enumi})}

\item the \emph{paired limit} $t' \gg t_{\parallel}, t_{\perp}$, which
we will discuss in detail in
Sec.~\ref{sect:infinitelystrongcorrelatedhops}.  In this limit, we
find SC correlations dominating at large distances (though, as for
hardcore bosons, CDW correlations inevitably dominate at short
distances).  Based on our numerical studies in
Sec.~\ref{sect:infinitelystrongcorrelatedhops}, the leading SC
correlation exponent appears to be universal, with a value of $\gamma
= \frac{1}{2}$, while the leading CDW correlation exponent $\beta$ is
nonuniversal.  In this limit, FL correlations are found to decay
exponentially. A staggered form of long-range CDW order also appears;

\item the \emph{two-leg limit} $t_{\perp} \ll t_{\parallel}$, $t' =
0$, which we will discuss in detail in
Sec.~\ref{sect:independentlegs}.  In this limit, the two legs of the
ladder are coupled only by infinite nearest-neighbor repulsion.  The
dominant correlations at large distances are those of a power-law CDW,
for which we find numerically to have what appears to be an universal
correlation exponent of $\beta = \frac{1}{2}$.  In this limit, the
leading SC correlation exponent was predicted analytically to be $\gamma =
2$, while FL correlations are found to decay exponentially;

\item the \emph{rung-fermion limit} $t_{\perp} \gg t_{\parallel}$, $t'
= 0$, which we will discuss in detail in
Sec.~\ref{sect:independentrungs}.  In this limit, the particles are
effectively localized onto the rungs of the ladder.  When the ladder
is quarter-filled, a true long-range CDW emerges in the two-fold
degenerate ground state.  Below quarter-filling, we find numerically
that the CDW power-law correlation dominate at large distances, with a
leading non-universal correlation exponent $\beta = \frac{1}{2} +
\frac{5}{2}\left(\frac{1}{2} - \Nbar_1\right)$.  The leading FL
correlation exponent was also found numerically to be non-universal,
with values going from $\alpha = 1$ to $\alpha = \frac{1}{4}$.  The SC
correlation exponent, on the other hand, was found numerically to be
universal, with value $\gamma = \frac{7}{4}$.

\end{enumerate}

\begin{figure}[htbp]
\includegraphics[scale=0.84]{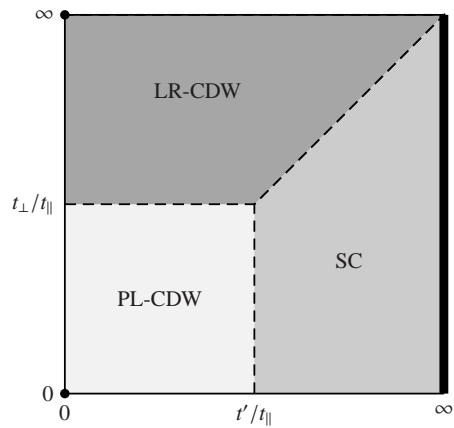}
\caption{The zeroth-order ground-state phase diagram of the ladder
model given by
Eq.~\eqref{eqn:anisotropicextendedHubbardcorrelatedhop}.  The three
limiting cases we can solve exactly are shown as the two dots (cases
(ii), power-law CDW (PL-CDW) and (iii), long-range CDW (LR-CDW)), and
the thick solid line (case (i), SC).}
\label{fig:roughphasediagram}
\end{figure}

To zeroth order (i.e.~without plunging into first-order perturbation
theory calculations), the ground-state phase diagram can be obtained
by interpolating between these three limiting cases.  There will be
three lines of quantum phase transitions or cross-overs, which at
quarter-filling, separate the long-range CDW (LR-CDW), power-law CDW
(PL-CDW), and SC phases.  At quarter-filling, we expect these three
lines of critical points or cross-overs to meet at a point on the
phase diagram.  If we have three lines of true critical points, this
point would be a quantum tricritical point.  We therefore end up with
a ground-state phase diagram which looks like that shown in Figure
\ref{fig:roughphasediagram}.

\subsection{The Paired Limit}
\label{sect:infinitelystrongcorrelatedhops}

In this subsection, we solve for the ground-state wave function, and
calculate various ground-state correlations in the paired limit $t'
\gg t_{\parallel}, t_{\perp}$.  In this limit, the Hamiltonian in
Eq.~\eqref{eqn:anisotropicextendedHubbardcorrelatedhop} simplifies to
\begin{equation}\label{eqn:HtpV}
\begin{aligned}
H_{t'V} &= -t'\sum_i\sum_j
\left(c_{i,j}^{\dagger} n_{i+1,j+1} c_{i,j+2} +
c_{i,j+2}^{\dagger} n_{i+1,j+1} c_{i,j}\right) \\
&\quad\ -\expandafter t'\sum_i\sum_j
\left(c_{i+1,j}^{\dagger} n_{i,j+1} c_{i+1,j+2} +
c_{i+1,j+2}^{\dagger} n_{i,j+1} c_{i+1,j}\right) \\
&\quad\ +\expandafter V\sum_i\sum_j n_{i,j} n_{i,j+1} +
V\sum_i\sum_j n_{i,j} n_{i+1,j}.
\end{aligned}
\end{equation}
In Sec.~\ref{sect:boundpairsdof}, we explain how pairs of spinless
fermions are bound by correlated hops in this limit, and the degrees
of freedom in the system become mobile bound pairs with infinite
nearest-neighbor repulsion.  These bound pairs come in two flavors,
determined by the specific arrangement of the two bound-pair particles
around a plaquette.  These flavors are conserved by correlated hops if
the length of the ladder is even, and hence the ladder ground state is
two-fold degenerate.  We then describe how these two degenerate ladder
ground states can be mapped to a excluded chain of hardcore bosons, then to
an ordinary chain of hardcore bosons, and finally to a chain of
noninteracting spinless fermions.

In Sec.~\ref{sect:tonkscorr}, we calculate the SC and CDW
correlations, using the intervening-particle expansion described in
Sec.~\ref{sect:nnegsexp}.  We then use a restricted-probability
argument in Sec.~\ref{sect:explainFLcorr} to show that FL correlations
decay exponentially with distance, governed by a density-dependent
correlation length, in this paired limit.  We find, as expected from
making the absolute correlated hopping amplitude $t'$ large, that SC
correlations dominate at large distances.

\subsubsection{Bound Pairs and Ground States}
\label{sect:boundpairsdof}

In the paired limit $t' \gg t_{\parallel}, t_{\perp}$, we solve for
the ground state of the simplified Hamiltonian given by
Eq.~\eqref{eqn:HtpV}, which admits only correlated hops.  Because of
this, isolated spinless fermions cannot hop at all; by contrast, a
pair occupying diagonal corners on a plaquette can perform correlated
hops.  Therefore, for an even number of spinless fermions,
ground-state configurations consist of well-defined bound pairs, which
are effectively bosons.  We say that a bound pair at $(1, j)$ and $(2,
j+1)$ has even (resp. odd) flavor if its two sites are even (resp.
odd) sites.  In this limit of $t'/t_{\parallel}, t'/t_{\perp} \to
\infty$, a particle on rung $j$ can only hop to rung $j \pm 2$.  This
moves the bound pair's center of mass by one lattice constant, without
changing its flavor.  The degrees of freedom in this limiting case
thus becomes bound pairs with definite flavors hopping along a
one-dimensional chain.\footnote{As we expect from having two flavors
of bound pairs, the many-bound-pair ground state is two-fold
degenerate for ladders of even length $L$ subject to periodic boundary
conditions.  For ladders of odd length $L$ subject to periodic
boundary conditions, the flavor of a bound pair changes as it goes
around the boundary of the ladder, and so the conserved quantum
numbers are not the even and odd flavors, but are instead the
symmetric and antisymmetric combinations of the two flavors.  This
mixing between even and odd flavors lifts the ground-state degeneracy,
giving a nondegenerate many-bound-pair ground state whose quantum
number is the antisymmetric combination of flavors.  In this paper we
consider only ladders of even length $L$, because we want to work with
ground states containing bound pairs with a definite flavor.}  We
write these hardcore boson operators in terms of the spinless fermion
operators as
\begin{equation}\label{eqn:evenboson}
B_{j,+}^{\dagger} = \begin{cases}
\cxd{1,j}\cxd{2,j+1}, & \text{$j$ odd}; \\
\cxd{1,j+1}\cxd{2,j}, & \text{$j$ even},
\end{cases}
\end{equation}
and
\begin{equation}\label{eqn:oddboson}
B_{j,-}^{\dagger} = \begin{cases}
\cxd{1,j+1}\cxd{2,j}, & \text{$j$ odd}; \\
\cxd{1,j}\cxd{2,j+1}, & \text{$j$ even}, \end{cases}
\end{equation}
where we order first with respect to the leg index, and then with
respect to the rung index of of the ladder.

Since bound pairs cannot move past each other along the chain, the
$P$-bound-pair Hilbert space breaks up into many independent sectors,
each with a fixed sequence of flavors.  The $P$-bound-pair problem in
one sector is therefore an independent problem from that of another
$P$-bound-pair sector.  The minimum energy in each sector can be very
crudely determined by treating the $P$-bound-pair problem as a
particle-in-a-box problem, where each bound pair is free to hop within
a `box' demarcated by its flanking bound pairs.  

\begin{figure}[htbp]
\includegraphics[scale=0.85]{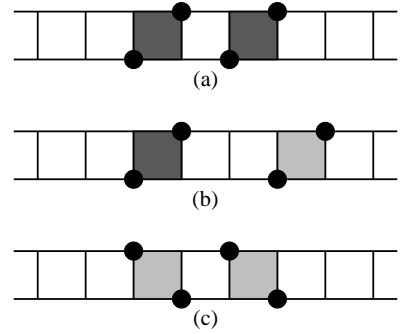}
\caption{The closest approach two bound pairs can make to each other, if
(a) they both have even flavors; (b) they have opposite flavors; and (c)
they both have odd flavors.}
\label{fig:closestapproach}
\end{figure}

As shown in Figure \ref{fig:closestapproach}, two bound pairs with the
same flavor can get within a separation $r = 2$ of each other, whereas
two bound pairs with different flavors can only achieve a closest
approach with separation $r = 3$.  Therefore, for a fixed separation
between flanking bound pairs, the kinetic energy of the `boxed' bound
pair is lowest when all three bound pairs have the same flavor.
Repeating this argument for all bound pairs, we realized therefore
that the two-fold degenerate ground state lies within the all-even and
all-odd sectors.  In these sectors, bound pairs cannot occupy
nearest-neighbor plaquettes, i.e. we are dealing with an excluded chain of
hardcore bosons.\footnote{More quantitatively, every sector maps to an
ordinary fermion chain, such that each change of flavor (between
successive pairs) diminishes the effective length $L'$ by one, thereby
increasing the particle density (and hence the energy of that sector's
ground state).}

The twofold degeneracy between all-even and all-odd sectors represents
a symmetry breaking with long-range order of a staggered CDW type (in
terms of fermion densities).  It may be viewed as breaking the
invariance under reflection about the ladder axis of the original
Hamiltonian as given in
Eq.~\eqref{eqn:anisotropicextendedHubbardcorrelatedhop}.\footnote{The
symmetry breaking has consequences for exact
diagonalizations~\cite{cheong07}.  Since we always have the same
number of spinless fermions on the two legs in this paired limit, we
expect reflection about the ladder axis to be an exact symmetry of the
ground states as well, as soon as $|t/t'|>0$ which permits a tiny
tunnel amplitude between the even and odd sectors in finite ladders.
The symmetrized ground states are $\tfrac{1}{\sqrt{2}}(\ket{\Psi_+}
\pm \ket{\Psi_-})$.} Thus the quantum-mechanical problem of a ladder
with density $\Nbar_2$ is mapped to the quantum-mechanical problem of
an excluded chain with density $\Nbar = \Nbar_2$.

\subsubsection{SC and CDW Correlations}
\label{sect:tonkscorr}

Three simple correlation functions, the FL, CDW, and SC correlations,
were computed for the excluded chain of hardcore bosons in
Sec.~\ref{sect:correlationsnnebosons}.  On the ladder model in the
paired limit, these correlations must be interpreted differently.  In
mapping the ladder model to the excluded chain, we replace a pair of
spinless fermion by a hardcore boson operator, i.e.~$\cxd{1,j}
\cxd{2,j+1} \to \Bxd{j}$.  Thus the FL correlation
$\braket{\Bxd{j}\Bx{j+r}}$ of hardcore bosons actually corresponds to
a SC correlation of the fermion model on the ladder.  Depending on
which of the two degenerate ladder ground states we are looking at,
the SC operators are
\begin{equation}\label{eqn:corrhopSCorderparam}
\begin{aligned}
\Delta_{j,g}^{\dagger} &= \tfrac{1}{\sqrt{2}}
(\cxd{1,j}\cxd{2,j+1} + \cxd{1,j+1}\cxd{2,j}), \\
\Delta_{j,u}^{\dagger} &= \tfrac{1}{\sqrt{2}} (-1)^j\,
(\cxd{1,j}\cxd{2,j+1} - \cxd{1,j+1}\cxd{2,j}),
\end{aligned}
\end{equation}
such that
\begin{equation}\label{eqn:nocrosscorr}
\begin{gathered}
\braket{\Delta_{j, g}^{\dagger} \Delta_{j+r, g}}_u = 
\braket{\Delta_{j, u}^{\dagger} \Delta_{j+r, u}}_g = 0, \\
\braket{\Delta_{j, g}^{\dagger} \Delta_{j+r, u}}_g = 
\braket{\Delta_{j, g}^{\dagger} \Delta_{j+r, u}}_u = 0.
\end{gathered}
\end{equation}
Because of Eq.~\eqref{eqn:nocrosscorr}, we shall drop the indices $g$
and $u$ from here on.  From Sec.~\ref{sect:correlationsnnebosons}, we
know that $\braket{\Delta_j^{\dagger} \Delta_{j+r}}$ decays with
separation $r$ asymptotically as the sum of a simple (leading) power
law and an $2k_F$-oscillatory (subleading) power law.  The leading
correlation exponent has been determined to be $\gamma_0 =
\frac{1}{2}$, while the subleading correlation exponent $\gamma_1$
cannot be reliably determined.

The simplest CDW correlations are 
\begin{equation}
\begin{gathered}
\braket{\cxd{1,j} \cx{1,j} \cxd{1,j+r} \cx{1,j+r}}, 
\braket{\cxd{1,j} \cx{1,j} \cxd{2,j+r} \cx{2,j+r}},\\
\braket{\cxd{2,j} \cx{2,j} \cxd{1,j+r} \cx{1,j+r}},
\braket{\cxd{2,j} \cx{2,j} \cxd{2,j+r} \cx{2,j+r}},
\end{gathered}
\end{equation}
which we call the CDW-$\sigma$ correlations.  These are not easy to
calculate, because they cannot be written simply in terms of the
expectations of hardcore boson operators.  In contrast, the CDW-$\pi$
correlations\footnote{The CDW-$\pi$ correlations $\braket{\Bxd{j}
\Bx{j} \Bxd{j+r} \Bx{j+r}}$ cannot be written as simple linear
combinations of eight-point functions because a term like
$\braket{\cxd{2,j} \cxd{1,j+1} \cx{1,j+1} \cx{2,j} \cxd{2,j+r+1}
\cxd{1,j+r} \cx{1,j+r} \cx{2,j+r+1}}$ will pick up contributions from
configurations that $\braket{\Bxd{j} \Bx{j} \Bxd{j+r} \Bx{j+r}}$ will
not.  This tells us that $\braket{\Bxd{j} \Bx{j} \Bxd{j+r} \Bx{j+r}}$
is some messy linear combination of eight-point, twelve-point,
sixteen-point, \dots, $4n$-point functions.}
\begin{equation}
\braket{\Bxd{j}\Bx{j}\Bxd{j+r}\Bx{j+r}} = \braket{N_j N_{j+r}}
\end{equation}
can be evaluated with the help of the intervening-particle expansion
in Eq.~\eqref{eqn:extendedhardcoreexpansion}.  This was done in
Sec.~\ref{sect:correlationsnnebosons}, where we found the subtracted
CDW-$\pi$ correlation $\braket{N_j N_{j+r}} - \braket{N_j}
\braket{N_{j+r}}$ decaying asymptotically with separation $r$ as a
simple power law $B_1\, r^{-\beta_1}\, \cos(2 k_F r + \theta_1)$, with
a nonuniversal leading correlation exponent $\beta_1 = \frac{1}{2} +
\frac{5}{2}\left(\frac{1}{2} - \Nbar_2\right)$, and a universal phase
shift of $\theta_1 = \pi/16$.

\subsubsection{FL Correlation: Explanation of Exponential Decay}
\label{sect:explainFLcorr}

Unlike the SC and CDW-$\pi$ correlations, the FL correlations cannot
be calculated easily in this paired limit, because the operators
involved cannot be written in terms of hardcore boson operators.
Nevertheless, we can still calculate it by making use of the fact that
this correlation is very close to being the probability of finding a
restricted class of configurations in the ground state.  We then make
use of the scaling form reported in Ref.~\onlinecite{cheong04b} to
calculate the probability analytically.  This idea is exploited again
in Sec.~\ref{sect:weakXY}

\begin{figure}[htbp]
\includegraphics[scale=0.85]{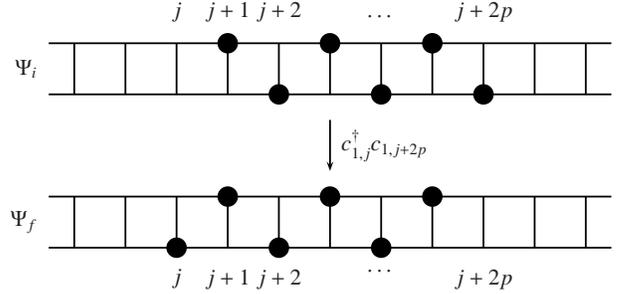}
\caption{Compact $p$-bound-pair cluster configurations making nonzero
contribution towards the expectation of the FL operator product
$\cxd{1,j}\cx{1,j+2p}$, which annihilates a spinless fermion on the
right end of the compact $p$-bound-pair cluster, and creates a
spinless fermion on the left end of the compact $p$-bound-pair
cluster.  $\Psi_i$ and $\Psi_f$ are the ground-state amplitudes of the
initial and final configurations respectively.}
\label{fig:compactclusterargument}
\end{figure}

In this paired limit, the ground state consists exclusively of a
superposition of bound pair configurations.  Therefore, if we
annihilate a spinless fermion on leg $i$, we must create another on
the same leg elsewhere, and thus the only nonzero FL correlations are
of the form $\braket{\cxd{i,j}\cx{i,j + r}}$.  In fact, to start with
a paired configuration and end up with another paired configuration,
after annihilating a spinless fermion at $j + r$ and creating a
spinless fermion at $j$, the initial and final configurations must
contain a compact cluster of pairs between rung $j$ and rung $j+r$, as
shown in Fig.~\ref{fig:compactclusterargument}.

Based on this compact cluster argument, we know that $\braket{
\cxd{i,j} \cx{i,j+r} } = 0$ when $r$ is odd.  When $r = 2p$ is even,
\begin{equation}
\braket{\cxd{i,j}\cx{i,j+2p}} = \sum_{(i, f)} \Psi_f^*\Psi_i
\end{equation}
receives contributions from all pairs of configurations with a compact
$p$-bound-pair cluster between the rungs $j$ and $j+2p$.  Clearly,
these products of amplitudes will depend on where the other bound
pairs are on the ladder.  However, if the ladder is not too close to
half-filling, we expect $\Psi_f \approx \Psi_i$, so that on an
infinite ladder, $\braket{\cxd{i,j} \cx{i,j+2p}}$ is very nearly the
probability of finding a compact $p$-bound-pair
cluster,\footnote{Technically, the correct thing to do is to compute
the $p$-particle sector of the cluster density matrix of a
$(p+1)$-site cluster, and look at the matrix element between a
configuration with an empty site at the left end of the cluster and a
configuration with an empty site at the right end of the cluster.
However, the relevant cluster density matrix is that of a system of
hardcore bosons.  While this hardcore-boson cluster density matrix
should be simply related to the noninteracting-spinless-fermion
cluster density matrix, this relation has not been worked out, for use
on this problem of finding FL correlations at large $r$ for the
bound-pair ground states on a two-legged ladder.}
\begin{equation}
\begin{aligned}
\braket{N_j N_{j+2} \cdots N_{j+2p}} &= 
\frac{\Nbar_2}{\nbar}\braket{n_j n_{j+1} \cdots n_{j+p}} \\
&= \frac{\Nbar_2}{\nbar}\det G_C(p),
\end{aligned}
\end{equation}
after using the relation \eqref{eqn:ON1On1} between excluded and ordinary
expectations, where $\nbar = \Nbar_2/(1 - \Nbar_2)$ is the density of
the ordinary chain.  Here $\det G_C(p)$ is the determinant of the
noninteracting-spinless-fermion cluster Green-function matrix $G_C(p)$
for a cluster of $p$ sites, which we can write as\cite{cheong04b}
\begin{equation}
\det G_C(p) = \prod_{l=1}^p \lambda_l = \prod_{l=1}^p 
\frac{1}{e^{\varphi_l} + 1},
\end{equation}
where $\lambda_l$ are the eigenvalues of the cluster Green-function matrix
$G_C(p)$, and $\varphi_l$ are the single-particle pseudo-energies of the
cluster density matrix $\rho_C$, for the cluster of $p$ sites in an
infinite chain of noninteracting spinless fermions.

For $p \gg 1$, $G_C(p)$ has approximately $\nbar p$ eigenvalues which
are almost one, and approximately $(1 - \nbar) p$ eigenvalues which
are almost zero, The determinant of $G_C(p)$ is thus determined
predominantly by the approximately $(1 - \nbar) p$ eigenvalues which
are almost zero.  For these $\lambda_l$, $e^{\varphi_l} \gg 1$, and
thus
\begin{equation}
\det G_C(p) \approx \prod_{\lambda_l \sim 0} e^{-\varphi_l} =
\exp\left(-\sum_{l_F}^{l_F + (1 - \nbar)p} \varphi_l\right),
\end{equation}
where $l_F$ is such that $\varphi_{l_F} = 0$.
Converting the sum into an integral, and using the approximate scaling
formula in Ref.~\onlinecite{cheong04b}, we find that
\begin{equation}
\det G_C(p) \approx \exp\left(-p\int_0^{1 - \nbar} 
f(\nbar, x)\,dx\right),
\end{equation}
i.e.~the probability of finding a compact $p$-bound-pair cluster decays
exponentially with $p$ in the limit of $p \gg 1$.  

With this simple compact cluster argument, we conclude that the ladder
FL correlation $\braket{\cxd{i,j} \cx{i,j+r}}$ decays exponentially
with separation $r$ as
\begin{equation}
\braket{\cxd{i,j}\cx{i,j+r}} \sim \exp\left[-r/\xi(\Nbar_2)\right],
\end{equation}
with a density-dependent correlation length
\begin{equation}\label{eqn:FLcorrelationlength}
\xi(\Nbar_2) = \frac{2}{\int_0^{1 - \nbar(\Nbar_2)} 
f(\nbar(\Nbar_2), x)\,dx},
\end{equation}
in the strong correlated hopping limit.  From
Ref.~\onlinecite{cheong04b} we know that the scaling function
$f(\nbar, x)$ depends only very weakly on $\nbar$, and thus, at very
low ladder densities $\Nbar_2 \to 0$, the correlation length
$\xi(\Nbar_2)$ attains its minimum value of
\begin{equation}
\xi(0) = \frac{2}{\int_0^1 f(0, x)\, dx},
\end{equation}
and the FL correlation $\braket{\cxd{i,j}\cx{i,j+r}}$ decays most rapidly
in this regime of $\Nbar_2 \to 0$.  This is expected physically, since
a long cluster of occupied sites is very unlikely to occur at very low
densities, with or without quantum correlations.

In the regime of $\Nbar_2 \to \frac{1}{2}$, we find $\nbar \to 1$, and
thus the correlation length $\xi(\Nbar)$ diverges according to
Eq.~\eqref{eqn:FLcorrelationlength}.  This diverging correlation
length tells us nothing about the amplitude of the FL correlation.
Indeed, when the ladder becomes half-filled, the two degenerate ground
states are inert bound-pair solids.  Each of the half-filled-ladder
ground-state wave functions consists of a single configuration whereby
all available plaquettes are occupied by a bound pair, and it is not
possible to annihilate a spinless fermion at the $(j+r)$th rung and
create another at the $j$th rung.  The FL correlation
$\braket{\cxd{i,j}\cx{i,j+r}}$ is thus strictly zero in this
half-filled-ladder limit.

\subsection{The Two-Leg Limit}
\label{sect:independentlegs}

This subsection concerns the ground state in the two-leg limit
$t_{\perp} \ll t_{\parallel}$, $t' = 0$.  Based on energetic
considerations, we argue in Section \ref{sect:nointerlegGS} that there
will be two degenerate ground states, within which successive spinless
fermions are on alternate legs of the ladder.  We call these the
\emph{staggered ground states}, and write their wave functions in
terms of the Fermi sea ground-state wave function with the help of a
\emph{staggered map} between ladder configurations and ordinary chain
configurations.  We then calculate various ground-state correlations
in Sec.~\ref{sect:weakXY}, Sec.~\ref{sect:weakXYCDW}, and
Sec.~\ref{sect:weakXYSC}, where we show that the non-vanishing FL
correlations decay exponentially with distance, governed by a
density-dependent correlation length, while the CDW and SC
correlations decay with distance as power laws.  We find in this
two-leg limit that the antisymmetric CDW correlation dominates at
large distances.

\subsubsection{Ground States}
\label{sect:nointerlegGS}

In the limit of $t_{\perp} \to 0$, each spinless fermion on the
two-legged ladder carries a permanent leg index, and thus the number
of spinless fermions $P_i$ on leg $i$ are good quantum numbers.
Furthermore, successive spinless fermions along the ladder cannot move
past each other, even if they are on different legs, because of the
infinite nearest-neighbor repulsion acting across the rungs.
Consequently, the Hilbert space of the $P$-spinless-fermion problem
breaks up into many independent sectors, each with a fixed sequence of
leg indices.  The $P$-spinless-fermion problem in one such sector is
therefore an independent problem from that of another
$P$-spinless-fermion sector.  Noting that the closest approach between
two particles on the same leg is $r = 2$, whereas that between two
particles on different legs is $r = 1$, we invoke the same
``particle-in-a-box'' argument used for the paired limit in
Sec.~\ref{sect:boundpairsdof} to find the ground state for $P$
spinless fermion on a ladder of even length $L$ to be in a
\emph{staggered} sector, where successive particles are on different
legs.  There are two such sectors in a ladder with open boundary
conditions, which we call sector 1 when the first fermion (from the
left) is on leg 1, or sector 2 when it is on leg 2.

Evidently this is a twofold symmetry breaking.  (The broken symmetry
is that of reflecting the configuration about the ladder axis, which
is a valid symmetry within the staggered sector.) This state has a
form of long range order, in that the flavor alternates; however, that
cannot be represented by any local order parameter, but only by a
``string'' order parameter~\cite{dennijs89}.

Let us write $\ket{\Psi_1}$ and $\ket{\Psi_2}$ for the ground
states in sectors 1 and 2, respectively.  
A staggered configuration of $P$
ladder spinless fermions in sector 1 can be mapped to a chain of $P$
noninteracting spinless fermions using the \emph{staggered map}
\begin{multline}\label{eqn:staggeredmap}
\cxd{1, j_1}\cxd{2, j_2}\cdots\cxd{1, j_{P-1}}\cxd{2, j_P}
\ket{0}_{\text{ladder}} \mapsto \\ 
\cxd{j_1}\cxd{j_2}\cdots\cxd{j_{P-1}}\cxd{j_P} \ket{0}_{\text{chain}}.
\end{multline}
Using the same formula, with an exchange of leg index 
$1\leftrightarrow 2$, 
configurations in sector 2 are similarly mapped~\footnote{
Whenever we explicitly discuss ground states, e.g. in studies
of exact diagonalization~\cite{cheong07}, it is appropriate
to make the ground states symmetric or antisymmetric under
reflection about the ladder axis,  namely
$\ket{\Psi_{\pm}} = (\ket{\Psi_1}\pm \ket{\Psi_2})/\sqrt{2}$.
If $|t_\perp/t|>0$, the sectors become connected with 
a tiny tunnel amplitude, in a finite system;
in this case only the symmetry-restored states
$\ket{\Psi_{\pm}}$ are actual eigenstates.}

Because the staggered map maps a ladder with density $\Nbar_2$ onto a
ordinary chain with density $\nbar = 2\Nbar_2$, we want a ladder observable
$O_{\text{ladder}}$ and its corresponding chain observable
$O_{\text{chain}}$ to be such that
\begin{equation}\label{eqn:OladderOchain}
\braket{O_{\text{ladder}}}_{\text{ladder}} = \tfrac{1}{2}
\braket{O_{\text{chain}}}_{\text{chain}}.
\end{equation}
This is analogous to Eq.~\eqref{eqn:ON1On1}, which we derived when we
map from a excluded chain to an ordinary chain.  We use the subscripts
`ladder' and `chain' just this once to distinguish between ladder and
chain expectations.  This notation is cumbersome, so we will not use
it again.  Whether an expectation is a ladder expectation or a chain
expectation should be clear from the context.  

\subsubsection{FL Correlations: Exponential Decay}
\label{sect:weakXY}

Having solved the staggered ground states in terms of the
one-dimensional Fermi sea, we calculate the FL, CDW, and SC
correlations.  There are four FL correlations at range $r$,
$\braket{\cxd{1,j}\cx{j+r}}$, $\braket{\cxd{1,j}\cx{2,j+r}}$,
$\braket{\cxd{2,j}\cx{1,j+r}}$ and $\braket{\cxd{2,j}\cx{2,j+r}}$.
From the staggered nature of $\ket{\Psi_{\pm}}$, we know that
\begin{equation}
\begin{aligned}
\braket{\cxd{1,j}\cx{1,j+r}} &= \braket{\cxd{2,j}\cx{2,j+r}}; \\
\braket{\cxd{1,j}\cx{2,j+r}} &= \braket{\cxd{2,j}\cx{1,j+r}}
\end{aligned}
\end{equation}
in both ground states.  The inter-leg FL correlations vanish, i.e.
\begin{equation}
\braket{\cxd{1,j}\cx{2,j+r}} = 0 = \braket{\cxd{2,j}\cx{1,j+r}},
\end{equation}
because annihilating a particle on one leg and creating a particle on
the other leg disrupts the stagger configuration.

\begin{figure}[htbp]
\centering
\includegraphics[scale=0.85]{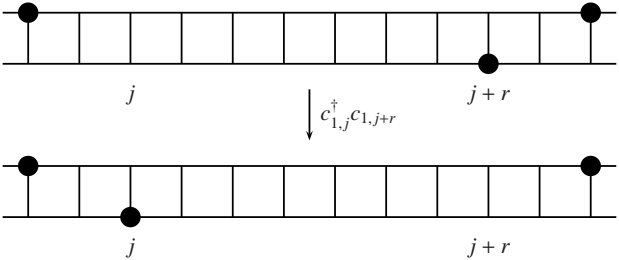}
\caption{Annihilation of a spinless fermion at site $(1, j+r)$, followed by
creation of a spinless fermion at site $(1, j)$, within a staggered
ground-state configuration leads to a staggered ground-state configuration,
when there are no intervening particles between rungs $j$ and $j + r$.}
\label{fig:staystagger}
\end{figure}

\begin{figure}[htbp]
\centering
\includegraphics[width=\linewidth,clip=true]{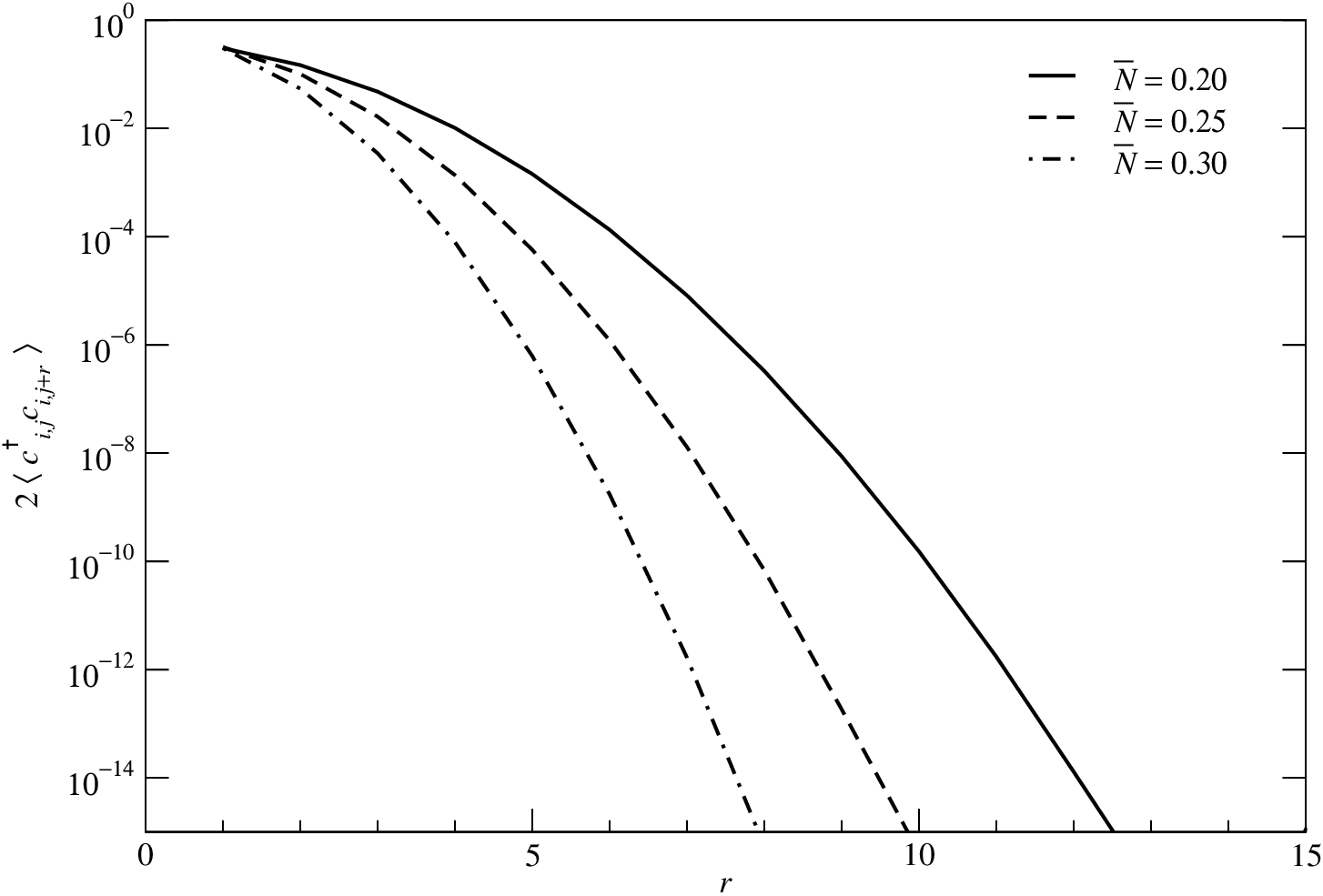}
\caption{The infinite-ladder FL correlations $\braket{\cxd{i,j}
\cx{i,j+r}}$, $i = 1, 2$, as a function of the separation $1 \leq r
\leq 15$ for ladder densities $\Nbar_2 = 0.20$, 0.25 and 0.30, in the
two-leg limit $t_{\perp}/t_{\parallel} \to 0$, $t' = 0$.}
\label{fig:ladderLinfXYweakFLr}
\end{figure}

The intra-leg FL correlation $\braket{\cxd{i, j}\cx{i, j+r}}$, which
is nonzero, receives contributions only from initial and final
staggered configurations in which there are no intervening particles
between rungs $j$ and $j+r$, for example, those shown in
Fig.~\ref{fig:staystagger}.  This tells us that
\begin{equation}\label{eqn:weakXYFL}
\braket{\cxd{i,j}\cx{i,j+r}} = \tfrac{1}{2}
\braket{\cxd{j}(\one - n_{j+1})\cdots(\one - n_{j+r-1})\cx{j+r}},
\end{equation}
when we map the ladder model to the chain model.  This correlation is
evaluated numerically, and shown in
Fig.~\ref{fig:ladderLinfXYweakFLr}, where we see the staggered
ground-state FL correlations decaying exponentially with separation
$r$.  This asymptotic behaviour can again be understood using a
constrained probabilities argument similar to that used in
Sec.~\ref{sect:tonkscorr}, except that instead of a compact cluster,
the relevant probability $P(r)$ is that of finding a gap at least $r$
in length within the one-dimensional Fermi-sea ground state.

Applying a restricted probability argument similar to the one outlined
in Sec.~\ref{sect:explainFLcorr}, we know this probability is simply
the zero-particle weight
\begin{equation}
P(r) = w_0 = \det(\one - G_C(r)),
\end{equation}
of the density matrix of a cluster of $r$ contiguous sites in the
chain of noninteracting spinless fermions.  For $r \gg 1$, the cluster
Green-function matrix $G_C(r)$ has approximately $(1 - \nbar)r$
eigenvalues which are almost zero, and $\nbar r$ eigenvalues which
are almost one.  The determinant of $\one - G_C(r)$ is thus
essentially determined by the approximately $\nbar r$ eigenvalues
which are almost one.   Using this fact, we calculate the asymptotic
form of $P(s)$ to be
\begin{equation}\label{eqn:gapprobability}
P(r) \approx \exp\left\{-r\int_0^{\nbar} f(1 - \nbar, x)\,dx\right\},
\end{equation}
where $f(\nbar, x)$ is the universaling scaling function identified in
Ref.~\onlinecite{cheong04b}.  Eq.~\eqref{eqn:gapprobability} explains
the observed exponential decay of $\braket{\cxd{i, j}\cx{i, j+r}}$ in
Fig.~\ref{fig:ladderLinfXYweakFLr}.  We note further that as $\nbar
\to 1$ (or equivalently, $\Nbar_2 \to \frac{1}{2}$), the FL
correlations decay fastest exponentially, whereas as $\nbar \to 0$
(equivalent to $\Nbar_2 \to 0$), the exponential decay is the slowest.
We expect these behaviours physically, because it is more likely to
find a long empty cluster when the density is low, and less likely to
find a long empty cluster when the ladder is closed to half-filled.

\subsubsection{CDW Correlations}
\label{sect:weakXYCDW}

Next, we calculate the CDW correlations, for which the four simplest
at separation $r$ are,
\begin{equation}
\begin{aligned}
\braket{\cxd{1,j}\cx{1,j}\cxd{1,j+r}\cx{1,j+r}} &= 
\braket{\cxd{2,j}\cx{2,j}\cxd{2,j+r}\cx{2,j+r}}, \\
\braket{\cxd{1,j}\cx{1,j}\cxd{2,j+r}\cx{2,j+r}} &=
\braket{\cxd{2,j}\cx{2,j}\cxd{1,j+r}\cx{1,j+r}}.
\end{aligned}
\end{equation}
Because of the staggered nature of the ground states, configurations
making nonzero contributions to $\braket{n_{i,j}n_{i,j+r}}$ are those
which map to noninteracting spinless fermion configurations in which
the sites $j$ and $j+r$ are occupied, with an odd number of
intervening particles between them.  Similarly, configurations making
nonzero contributions to $\braket{n_{i,j} n_{i',j+r}}, i \neq i'$, are
those which map to noninteracting spinless fermions in which the sites
$j$ and $j+r$ are occupied, with an even number of intervening
particles between them.  

Defining the density operators
\begin{equation}\label{eqn:weakXYnpm}
n_{\pm, j} \equiv n_{1,j} \pm n_{2, j}
\end{equation}
which are symmetric and antisymmetric with respect to reflect along
the ladder axis, we find that
\begin{equation}
\braket{n_{+, j} n_{-, j+r}} = 0 = \braket{n_{-,j} n_{+, j+r}},
\end{equation}
and
\begin{equation}\label{eqn:Sigmaplus}
\braket{n_{+, j}n_{+, j+r}} = \braket{n_jn_{j+r}} = \Sigma_+(r),
\end{equation}
which we call the \emph{CDW$+$ correlation}.  This is identical to the
CDW correlation of the one-dimensional Fermi sea, which we know decays
as an oscillatory power law
\begin{equation}
\braket{n_{+,j} n_{+, j+r}} - \braket{n_{+,j}}\braket{n_{+,j+r}}
\sim r^{-2}\, \cos(2k_F r).
\end{equation}

\begin{figure}[htbp]
\centering
\includegraphics[width=\linewidth,clip=true]{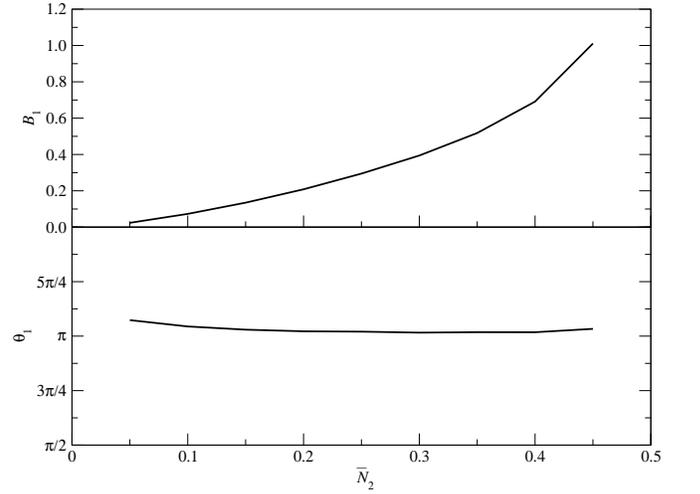}
\caption{Plot of the fitted amplitude $B_1$ (top) and fitted phase
shift $\theta_1$ (bottom) of the leading oscillatory power-law decay,
as functions of the ladder density $\Nbar_2$, for the CDW$-$
correlation in the staggered ground state of the ladder model, in the
two-leg limit $t_{\perp}/t_{\parallel} \to 0$, $t' = 0$.}
\label{figure:weakXYcdwmfits}
\end{figure}

There is also the \emph{CDW$-$ correlation}, 
\begin{equation}\label{eqn:nmnm}
\braket{n_{-, j}n_{-, j+r}} = 2\left(\braket{n_{1,j}n_{1,j+r}} - 
\braket{n_{1,j}n_{2,j+r}}\right) = \Sigma_-(r),
\end{equation}
associated with $n_{-,j}$.  This is identical to the subtracted CDW$-$
correlation, since $\braket{n_{-,j}} = \braket{n_{1, j} - n_{2, j}} =
0$.  Evaluating this expectation numerically, we find that at all
ladder densities $\Nbar_2$, $\Sigma_-(\Nbar_2, r)$ oscillates
about a zero average with wave vector $2k_F$, and a decaying
amplitude.  A preliminary unrestricted nonlinear curve fitting to the
asymptotic form
\begin{equation}\label{eqn:CDWAakphi}
\Sigma_-(\Nbar_2, r) = B_0\, r^{-\beta_0} + B_1\, r^{-\beta_1}\,
\cos(2 k_F r + \theta_1),
\end{equation}
where $B_1\, r^{-\beta_1}\, \cos(2 k_F r + \theta_1)$ is the leading
asymptotic behaviour, and $B_0\, r^{-\beta_0}$ is a correction term,
suggests that the leading correlation exponent may actually be
universal, taking on the value $\beta_1 = \frac{1}{2}$.  Further
nonlinear curve fitting, restricting $\beta_1 = \frac{1}{2}$, tells us
that only the parameters $B_1$ and $\theta_1$ of the leading
asymptotic term can be reliably determined.  These are shown in
Fig.~\ref{figure:weakXYcdwmfits}.  From this restricted curve fit, it
appears that the phase shift might also be universal, taking on value
$\theta_1 = \pi$.

\subsubsection{SC Correlations}
\label{sect:weakXYSC}

The simplest SC correlations at separation $r$ are
\begin{equation}
\begin{gathered}
\braket{\cxd{1,j}\cxd{2,j+1}\cx{1,j+r}\cx{2,j+r+1}} = 
\braket{\cxd{2,j}\cxd{1,j+1}\cx{2,j+r}\cx{1,j+r+1}}, \\
\braket{\cxd{1,j}\cxd{2,j+1}\cx{2,j+r}\cx{1,j+r+1}} =
\braket{\cxd{2,j}\cxd{1,j+1}\cx{1,j+r}\cx{2,j+r+1}}.
\end{gathered}
\end{equation}
Correlations of the type $\braket{\cxd{i,j} \cxd{i',j+1} \cx{i,j+r}
\cx{i',j+r+1}}$ receive nonzero contributions from configurations
containing an even number of intervening particles between rungs $j+1$
and $j+r$, whereas correlations of the type $\braket{\cxd{i,j}
\cxd{i',j+1} \cx{i',j+r} \cx{i,j+r+1}}$ receive nonzero contributions
from configurations containing an odd number of intervening particles
between rungs $j+1$ and $j+r$.  Defining the paired operators
\begin{equation}
\Delta_{\pm, j}^{\dagger} = \tfrac{1}{\sqrt{2}}\left(\cxd{1,j}\cxd{2,j+1} 
\pm \cxd{1,j+1}\cxd{2,j}\right),
\end{equation}
which are symmetric and antisymmetric with respect to reflection about
the ladder axis, we find that
\begin{equation}
\braket{\Delta_{+,j}^{\dagger}\Delta_{-,j+r}} = 0 =
\braket{\Delta_{-,j}^{\dagger}\Delta_{+,j+r}},
\end{equation}
and that the SC$+$ correlation
\begin{equation}
\braket{\Delta_{+,j}^{\dagger}\Delta_{+,j+r}} =
\braket{\cxd{j}\cxd{j+1}\cx{j+r}\cx{j+r+1}} = \Pi_+(r) \sim r^{-2}
\end{equation}
at large separations.

\begin{figure}[htbp]
\centering
\includegraphics[width=\linewidth,clip=true]{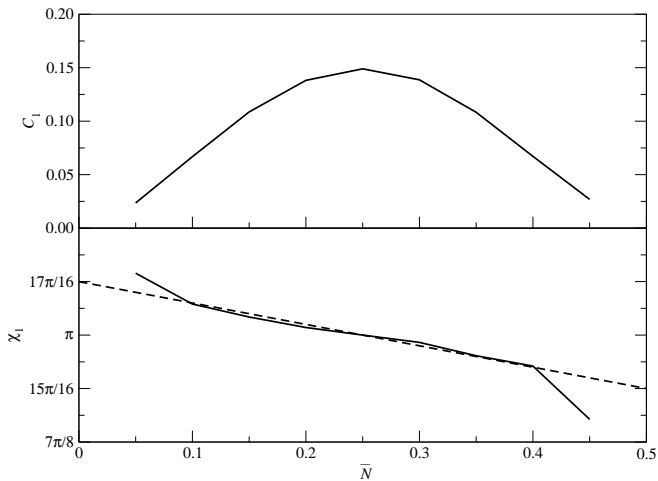}
\caption{Plot of the fitted amplitude $C_1$ (top) and fitted phase
shift $\chi_1$ (bottom) of the leading oscillatory power-law decay, as
functions of the ladder density $\Nbar_2$, for the SC$-$ correlation
in the staggered ground state of the ladder model, in the two-leg
limit $t_{\perp}/t_{\parallel} \to 0$, $t' = 0$.}
\label{figure:weakXYscmfits}
\end{figure}

The SC$-$ correlation
\begin{equation}
\braket{\Delta_{-,j}^{\dagger}\Delta_{-,j+r}} = \Pi_-(r)
\end{equation}
must be evaluated numerically.  We find that, just like $\Sigma_-(r)$,
$\Pi_-(r)$ oscillates about a zero average with wave vector $2k_F$,
and a rapidly decaying amplitude.  To improve the quality of the
nonlinear curve fitting, we fit $r^2 \Pi_-(r)$ to the asymptotic form
\begin{equation}
r^2 \Pi_-(r) = C_0\, r^{2 - \beta_0} + C_1\, r^{2 - \beta_1}\, 
\cos(2 k_F r + \chi_1),
\end{equation}
where $C_1\, r^{2 - \beta_1}\, \cos(2 k_F r + \chi_1)$ is the leading
asymptotic behavior, while $C_0\, r^{2 - \beta_0}$ is a correction
term.  A preliminary unrestricted fit suggests that the leading
correlation exponent is universal, and takes on value $\beta_1 =
\frac{5}{2}$.  Further restricted nonlinear curve fitting tells us
that only the parameters $C_1$ and $\chi_1$ can be reliably
determined.  These are shown in Fig.~\ref{figure:weakXYscmfits},
where we see that the amplitude $C_1$ exhibits symmetry about quarter
filling, which is a kind of particle-hole symmetry, and that the phase
shift $\chi_1 = \pi\left[1 + \frac{1}{4}\left(\frac{1}{4} -
\Nbar_2\right)\right]$ is non-universal, but depends linearly on the
density $\Nbar_2$.

\subsection{The Rung-Fermion Limit}
\label{sect:independentrungs}

In this subsection, we look at the rung-fermion limit 
$t_{\perp} \gg t_{\parallel}$, $t' = 0$.  We argue in Section
\ref{sect:stronginterlegGS} that in this limit, each spinless fermion
spends most of its time hopping back and forth along the rung it is
on, and only very rarely hops along the legs to an adjacent rung.
Therefore, each spinless fermion will be in a quantum state very close
to the symmetric eigenstate of one rung, and we can think of the
ladder of spinless fermions with density $\Nbar_2$ in this limit as
essentially an excluded chain of fermions with density $\Nbar =
2\Nbar_2$.  For $\Nbar_2 < \frac{1}{4}$, the ground state of this
excluded chain of rung-fermions has been solved in
Sec.~\ref{sect:nnetonni}.  The FL, CDW, and SC correlations have also
been calculated in Sec.~\ref{sect:correlationsnnebosons}, so we will
not repeat them here.

At $\Nbar_2 = \frac{1}{4}$, the ground state is a `dynamic solid'
phase, in which rung-fermions occupy either all the even rungs, or all
the odd rungs, and cannot hop along the legs to adjacent rungs because
of the infinite nearest-neighbor repulsion between them.  For $\Nbar_2
> \frac{1}{4}$, we describe in Sec.~\ref{sect:phasesep} how the system
will phase separate into a high-density inert solid phase, in which
spinless fermions cannot hop at all, and the lower-density `dynamic
solid' phase.  In this phase separation regime, the FL, CDW, and SC
correlations cannot be calculated.

\subsubsection{Ground States}
\label{sect:stronginterlegGS}

In the limit of $t_{\parallel}/t_{\perp} \to 0$, a spinless fermion
spends most of its time hopping back and forth along a rung, and only
very rarely hops along the leg to an adjacent rung, where it will
spend a lot of time hopping back and forth, before hopping along the
leg again.  Because of this long dwell time on a rung, the spinless
fermion is very nearly in the rung ground state
\begin{equation}
\ket{+, j} = \tfrac{1}{\sqrt{2}}\left(\cxd{1,j} + \cxd{2,j}\right)\ket{0} =
\Cxd{j}\ket{0}.
\end{equation}
Let us call a spinless fermion in the rung ground state a \emph{rung
fermion} in short.  Rung-fermions inherit the infinite
nearest-neighbor repulsion of the bare spinless fermions, and
therefore two rung-fermions in adjacent rungs experience infinite
nearest-neighbor repulsion as well.  With this insight, we find that
the full many-body problem of spinless fermions with infinite
nearest-neighbor repulsion on the two-legged ladder with density
$\Nbar_2$ reduces to the problem of an excluded chain with density
$\Nbar = 2\Nbar_2$ of spinless rung-fermions.

The latter problem was solved in Sec.~\ref{sect:maptech} and
Sec.~\ref{sect:correlationsnnebosons} for excluded chain densities $\Nbar
< \frac{1}{2}$.  In the special case of quarter-filling on the ladder,
$\Nbar_2 = \frac{1}{4}$, spinless fermions occupy alternate rungs.
These are free to hop along the rungs that they
reside on, but cannot hop along the legs, for non-vanishing values of
$t_{\parallel}/t_{\perp}$.  Even virtual processes in which a spinless
fermion on rung $j$ hops along the leg to an adjacent rung and back
are essentially forbidden by the infinite nearest-neighbor repulsion,
because such virtual processes, which has a time scale of
$O(1/t_{\parallel})$, would not be complete when the spinless fermion
on the next-nearest-neighbor rung hops across the rung, which occurs
on a time scale of $O(1/t_{\perp})$.  Virtual processes such as these
only become energetically feasible when the two time scales become
comparable, i.e.~when $t_{\parallel} \lesssim t_{\perp}$.  Therefore,
over a wide range of anisotropies $t_{\parallel}/t_{\perp}$, the
spinless fermions in the quarter-filled ladder with $t' = 0$ can hop
back and forth along the rungs they are on, but cannot hop to the
neighboring rungs.  This gives rise to a symmetry breaking, where the
spinless fermions are either all on the even rungs, or they are all on
the odd rungs.  Because translational symmetry along the ladder axis
is broken in the quarter-filled ladder ground states, we think of
these as `dynamic solids', since the constituent spinless fermions are
constantly hopping back and forth along the rungs.  In this limit, the
only non-vanishing correlation is the rung-fermion CDW correlation
\begin{equation} 
\braket{N_j N_{j+r}} = \begin{cases}
\frac{1}{2}, & \text{$r$ even}; \\ 
0, & \text{$r$ odd}, \end{cases}
\end{equation} 
i.e.~there is true long-range order in the quarter-filled ladder
ground state in the limit of $t_{\perp} \gg t_{\parallel}$, $t' = 0$.

\subsubsection{Phase Separation}
\label{sect:phasesep}

\begin{figure}[htbp]
\centering
\includegraphics[scale=0.75]{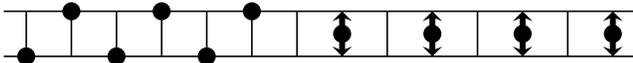}
\caption{Phase separation of a greater-than-quarter-filled ladder of
spinless fer\-mi\-ons with infinite nearest-neighbor repulsion into a
high-density inert solid phase (immobile spinless fermions) with
$\Nbar_2 = \frac{1}{2}$, and a low-density fluid phase (mobile
spinless fermions shown with arrows) with $\Nbar_2 = \frac{1}{4}$, in
the rung-fermion limit $t_{\perp} \gg t_{\parallel}$, $t' = 0$.}
\label{fig:stronginterlegphaseseparation}
\end{figure}

In this rung-fermion limit $t_{\perp} \gg t_{\parallel}$, $t' = 0$,
the system phase separates for ladder densities $\Nbar_2 >
\frac{1}{4}$.  As shown in
Fig.~\ref{fig:stronginterlegphaseseparation}, when the ladder is above
quarter-filling, some of the spinless fermions will go into a
high-density inert solid phase with density $\Nbar_2 = \frac{1}{2}$,
where spinless fermions are arranged in a staggered array, and
therefore cannot hop at all.  These spinless fermions contribute
nothing to the ground-state energy.  If $t_{\parallel}$ is comparable
to $t_{\perp}$, the rest of the spinless fermions will go into a fluid
phase, whose density is $\Nbar_2 < \frac{1}{4}$.  These spinless
fermions are free to hop back and forth on the rungs they are on, and
occasionally to the neighboring rungs, when permitted by
nearest-neighbor exclusion.  These contribute a density-dependent
total kinetic energy to the ground-state energy.  The ground-state
composition depends on whether the kinetic energy gained per particle,
by removing a spinless fermion from the solid phase and adding it to
the fluid phase, outweighs the decrease in kinetic energy per particle
that results from the fluid becoming more congested.

When $t_{\perp}$ becomes large compared to $t_{\parallel}$, which is
the limit we are interested in, it becomes energetically favorable,
always, to remove one spinless fermion from the inert solid phase, and
add it to the fluid phase, if its density is $\Nbar_2 < \frac{1}{4}$.
This is because the kinetic energy penalty to make the fluid becoming
more congested, which is of $O(t_{\parallel})$, is more than
compensated for by the kinetic energy gain of $t_{\perp}$ for an extra
spinless fermion freed to hop back and forth along a rung.  Iterating
this argument, we find then that, for $t_{\perp} \gg t_{\parallel}$,
and the overall density $\Nbar_2 > \frac{1}{4}$, the system will phase
separate into an inert solid phase with density $\Nbar_2 =
\frac{1}{2}$, and a dynamic solid phase with density $\Nbar_2 =
\frac{1}{4}$.  For example, if the overall density is $\Nbar_2 =
\frac{1}{3} > \frac{1}{4}$, we will find that $\frac{1}{3}$ of the
total number of spinless fermions will be in the inert solid phase,
while the other $\frac{2}{3}$ of the total number of spinless fermions
will be in the dynamic solid phase.

\section{Summary and Discussions}
\label{sect:conclusions}

In this paper, we established a one-to-one correspondence between
$P$-particle configurations on the excluded chain and $P$-particle
configurations on the ordinary chain using the right-exclusion map.
We then showed that the Hamiltonian matrices of the models given in
Eq.~\eqref{eqn:HBHC} and Eq.~\eqref{eqn:HbHc} are identical, therefore
solving for the ground states of the former in terms of those of the
latter.  These results were obtained for finite chains subject to open
boundary conditions, but continues to hold for infinite chains.

Based on this one-to-one correspondence between ground states, we
showed that the ground-state expectation $\braket{O}$ of an excluded
chain observable $O$ can be evaluated using Eq.~\eqref{eqn:ON1On1} in
terms of the ground-state expectation $\braket{O'}$ of a carefully
chosen corresponding observable $O'$ on the ordinary chain.  We then
developed the method of intervening-particle expansion, to write the
ground-state expectation $\braket{O_j O_{j+r}}$ of a product of local
excluded chain operators $O_j$ and $O_{j+r}$, first as a sum over
excluded chain expectations $\braket{O_j O_p O_{j+r}}$ conditioned on
the occupations of the sites between $j$ and $j+r$, and then as a sum
over the corresponding ordinary chain expectations $\braket{O'_j O'_p
O'_{j+r-p}}$. 

Using these analytical results from Sec.~\ref{sect:maptech}, we
calculated the FL, CDW, and SC correlations of the excluded chains of
hardcore bosons and spinless fermions in
Sec.~\ref{sect:correlationsnnebosons}.  Based on nonlinear curve fits
of the numerically evaluated correlations, to reasonable asymptotic
forms, we find all three types of correlations decaying with
separation $r$ as power laws, for hardcore bosons as well as for
spinless fermions.  More interestingly, we find for both hardcore
bosons and spinless fermions a universal exponent $\gamma_1 =
\frac{7}{4}$ for the oscillatory power-law decay of the SC
correlation, but a non-universal, density-dependent, exponent $\beta_1
= \frac{1}{2} + \frac{5}{2} (\frac{1}{2} - \Nbar)$ for the oscillatory
power-law decay of the CDW correlation.  Also, the leading asymptotic
behaviour for the hardcore boson FL correlation was found to a
non-oscillating power-law decay with universal exponent $\alpha_0 =
\frac{1}{2}$, while that for the spinless fermion FL correlation was
found to be oscillations in a power-law envelope, with a non-universal
exponent that approaches $\alpha_1 = 1$ as $\Nbar \to 0$, and
$\alpha_1 = \frac{1}{4}$ as $\Nbar \to \frac{1}{2}$.

We then analyzed our spinless-fermion ladder model,
Eq.~\eqref{eqn:anisotropicextendedHubbardcorrelatedhop}, in
Sec.~\ref{sect:laddermodels}.  This ladder model can be solved exactly
in three limiting cases: (i) the paired limit $t' \gg t_{\parallel},
t_{\perp}$; (ii) the two-leg limit $t_{\perp} \ll t_{\parallel}, t' =
0$; and (iii) the rung-fermion limit $t_{\perp} \gg t_{\parallel}, t'
= 0$.  In the paired limit, which we solved in
Sec.~\ref{sect:infinitelystrongcorrelatedhops}, spinless fermions form
correlated-hopping bound pairs, and so the ladder model can be mapped
to the excluded chain of hardcore bosons.  The ground state of this latter
model was solved exactly in Sec.~\ref{sect:maptech} and its
ground-state correlations calculated in
Sec.~\ref{sect:correlationsnnebosons}.  By reinterpreting the excluded
chain correlations in ladder terms, we realized that ladder SC
correlations dominates at large distances over ladder CDW
correlations, both of which decay as power laws with separation, with
leading exponents $\gamma_0 = \frac{1}{2}$ and $\beta_1 = \frac{1}{2}
+ \frac{5}{2}(\frac{1}{2} - \Nbar_2)$ respectively, $\Nbar_2$ being
the ladder density.  We also showed, using a restricted probabilities
argument, that ladder FL correlations decay exponentially with
separation, with a density-dependent correlation length.

Next, in the two-leg limit, which we solved in
Sec.~\ref{sect:independentlegs}, we argued based on a
``particle-in-a-box'' picture that successive spinless fermions in the
two-fold degenerate \emph{staggered ground states} occupy different
legs of the ladder.  We write these ground states exactly in terms of
the one-dimensional Fermi sea in Sec.~\ref{sect:nointerlegGS}, before
calculating correlations in Sec.~\ref{sect:weakXY}.  We found, using a
different restricted probabilities argument, that FL correlations
decay exponentially with separation, with a density-dependent
correlation length.  CDW and SC correlations symmetric (antisymmetric)
with respect to a reflection about the ladder axis decay as power
laws, with universal leading exponents $\beta_1 = 2 (\frac{1}{2})$ and
$\gamma_1 = 2 (\frac{5}{2})$ respectively.

Finally, in the rung-fermion limit, we mapped the ladder model to an
excluded chain of spinless fermions in
Sec.~\ref{sect:independentrungs}.  Since we have already solved this
latter model in Sec.~\ref{sect:maptech} and calculated its
ground-state correlations in Sec.~\ref{sect:correlationsnnebosons}
below half-filling (which corresponds to quarter-filling on the
ladder), we discussed the phase separations that occurs on ladders
with greater than quarter filling in Sec.~\ref{sect:stronginterlegGS}.
Correlation exponents obtained for the three limiting cases of our
ladder model Eq.~\eqref{eqn:anisotropicextendedHubbardcorrelatedhop},
as well as those for the excluded chains of hardcore bosons and
spinless fermions, are summarized in Table
\ref{table:summaryofexponents}.

\begin{table}
\centering
\caption{A summary of the leading correlation exponents and wave
vectors of various correlation functions that decay as power laws in
the (i) paired limit $t' \gg t_{\parallel}, t_{\perp}$; (ii) two-leg
limit $t_{\perp} \ll t_{\parallel}$, $t' = 0$; and (iii) rung-fermion
limit $t_{\perp} \gg t_{\parallel}$, $t' = 0$.  The wave vector $k$ of
the leading terms in the correlation functions are reported in terms
of $k_F = \pi\Nbar_1$, where $0 \leq \Nbar_1 \leq \frac{1}{2}$ is the
excluded chain density.  The suffixes $\pi$ and $\pm$ indicate further
symmetries possible in the ladder model.}
\label{table:summaryofexponents}
\vskip \baselineskip
\begin{tabular}{cccc}\hline
\parbox[c]{.28\linewidth}{\centering model} & 
\parbox[c]{.22\linewidth}{\centering correlation function} & 
\parbox[c]{.22\linewidth}{\centering correlation exponent} & 
\parbox[c]{.20\linewidth}{\centering wave vector} \\ \hline
hardcore boson & FL & $\frac{1}{2}$ & 0 \\ \cline{2-4}
 & CDW & $\frac{1}{2} + \frac{5}{2}\left(\frac{1}{2} - \Nbar_1\right)$
 & $2k_F$ \\ \cline{2-4}
 & SC & $\frac{7}{4}$ & 0 \\ \hline
spinless fermion & FL & $1 \to \frac{1}{4}$ & $k_F$ \\ \cline{2-4}
 & CDW & $\frac{1}{2} + \frac{5}{2}\left(\frac{1}{2} - \Nbar_1\right)$
 & $2k_F$ \\ \cline{2-4}
 & SC & $\frac{7}{4}$ & 0 \\ \hline
$t' \gg t_{\parallel}, t_{\perp}$ & CDW-$\pi$ & $\frac{1}{2} +
\frac{5}{2}\left(\frac{1}{2} - \Nbar_1\right)$ & $2k_F$ \\ 
 & & 2 & 0 \\ \cline{2-4}
 & SC & $\frac{1}{2}$ & 0 \\
 & & $\frac{3}{2} \to \frac{1}{2}$ & $2k_F$ \\ \hline
$t_{\perp} \ll t_{\parallel}$, $t' = 0$ & CDW$+$ & 2 & 0 \\
 & & 2 & $2k_F$ \\ \cline{2-4}
 & CDW$-$ & $\frac{1}{2}$ & $2k_F$ \\
 & & 2 & 0 \\ \cline{2-4}
 & SC$+$ & 2 & 0 \\
 & & 2 & $2k_F$ \\ \cline{2-4}
 & SC$-$ & $\frac{5}{2}$ & $2k_F$ \\
 & & 4 & 0 \\ \hline
$t_{\perp} \gg t_{\parallel}$, $t' = 0$ 
 & FL & $1 \to \frac{1}{4}$ & $k_F$ \\ \cline{2-4}
 & CDW & $\frac{1}{2} + \frac{5}{2}
 \left(\frac{1}{2} - \Nbar_1\right)$ & $2k_F$ \\
 & & 2 & 0 \\ \cline{2-4}
 & SC & $\frac{7}{4}$ & 0 \\ \hline
\end{tabular}
\end{table}

In this study, we find the emergence of surprising universal
correlation exponents.  In the Luttinger liquid paradigm, all
correlation exponents can be written in terms of the exponents
\cite{solyom79,voit92,voit95} 
\begin{equation}\label{eqn:gammaparameter}
\begin{aligned}
\gamma_{\rho} &= \tfrac{1}{8}(K_{\rho} + K_{\rho}^{-1} - 2), \\
\gamma_{\sigma} &= \tfrac{1}{8}(K_{\sigma} + K_{\sigma}^{-1} - 2) 
\end{aligned}
\end{equation}
appearing in the quantum-mechanical propagator, also called the
(equal-time) two-point function
\begin{equation}
\begin{aligned}
G(r) &\sim A_1\, r^{-\alpha} \cos k_F r = 
A_1\, r^{-[1 + 2 (\gamma_{\rho} + \gamma_{\sigma})]} \cos k_F r \\
&= A_1\, r^{-\frac{1}{4}\left[(K_{\rho} + K_{\rho}^{-1}) +
(K_{\sigma} + K_{\sigma}^{-1})\right]} \cos k_F r.
\end{aligned}
\end{equation}
The parameters $K_{\rho}$ and $K_{\sigma}$ depend generically on the
filling fraction and the interaction strength, and thus all
correlation exponents are non-universal.  In particular, various
theoretical approaches (see review by S\'olyom \cite{solyom79}) tell
us that the charge density waves (CDW), spin density wave (SDW),
singlet superconductivity (SSC) and triplet superconductivity (TSC)
correlations decay as power laws
\begin{subequations}\label{eqn:expKrhoKsigma}
\allowdisplaybreaks
\begin{align}
\braket{n(0)n(r)} &\sim \frac{K_{\rho}}{\pi^2 r^2} +
B_2 r^{-K_{\rho}-K_{\sigma}} \cos 2 k_F r + {} \notag \\
&\quad\ B_4 r^{-4 K_{\rho}} \cos 4 k_F r, \\
\braket{\sigma_x(0)\sigma_x(r)} &= 
\braket{\sigma_y(0)\sigma_y(r)} \notag \\
&\sim \frac{D_{0,xy}}{r^2} + 
D_{2,xy} r^{-K_{\rho} - K_{\sigma}^{-1}} \cos 2 k_F r, \\
\braket{\sigma_z(0)\sigma_z(r)} &\sim \frac{D_{0,z}}{r^2} + 
D_{2,z} r^{-K_{\rho} - K_{\sigma}} \cos 2 k_F r, \\
\braket{\Delta_{0,0}^{\dagger}(0)\Delta_{0,0}(r)} &=  
\braket{\Delta_{1,0}^{\dagger}(0)\Delta_{1,0}(r)} 
\sim  C_0 r^{-K_{\rho}^{-1} - K_{\sigma}}, \\
\braket{\Delta_{1,\pm 1}^{\dagger}(0)\Delta_{1,\pm 1}(r)} &\sim 
C'_0 r^{-K_{\rho}^{-1} - K_{\sigma}^{-1}}
\end{align}
\end{subequations}
in a Tomonaga-Luttinger liquid.

When the chain of interacting spinfull fermions is spin-rotation
invariant (for example, in the absence of an external magnetic field),
the spin stiffness constant must be $K_{\sigma} = 1$, and the
ground-state properties become completely determined by the single
nontrivial Luttinger parameter $K_{\rho}$.  The spinfull power laws
thus become
\begin{subequations}
\allowdisplaybreaks
\begin{align}
G(r) &\sim A_1 r^{-\frac{1}{4}\left(K_{\rho} + K_{\rho}^{-1} + 2\right)}
\cos k_F r, \\
\braket{n(0)n(r)} &\sim \frac{K_{\rho}}{\pi r^2} +
B_2 r^{-K_{\rho}-1} \cos 2 k_F r + {} \notag \\
&\quad\ B_4 r^{-4 K_{\rho}} \cos 4 k_F r, \\
\braket{\boldsymbol{\sigma}(0)\cdot\boldsymbol{\sigma}(r)} &\sim 
\frac{1}{\pi r^2} + D_2 r^{-K_{\rho} - 1} \cos 2 k_F r, \\
\braket{\Delta_{0}^{\dagger}(0)\Delta_{0}(r)} &=  
\braket{\Delta_{1}^{\dagger}(0)\Delta_{1}(r)} \sim  
C_0\, r^{-K_{\rho}^{-1} - 1}.
\end{align}
\end{subequations}
For spinless fermions, there is only one independent stiffness
constant $K_{\rho} = K_{\sigma} = K$ \cite{luther75,voit95}, so that 
the spinfull power laws which have proper spinless analogs are
\begin{subequations}
\allowdisplaybreaks
\begin{align}
G(r) &\sim A_1 r^{-\frac{1}{2}\left(K + K^{-1}\right)} \cos k_F r, \\
\braket{n(0)n(r)} &\sim \frac{K}{\pi r^2} + 
B_2 r^{-2K} \cos 2 k_F r + {} \notag \\
&\quad\ B_4 r^{-4 K} \cos 4 k_F r.
\end{align}
\end{subequations}

In the Luttinger liquid paradigm, universal correlation exponents only
arise in the special case of the Fermi liquid, where we have $K_{\rho}
= 1$.  Consequently, the two-point function decays as $G(r) \sim
r^{-1}$, while the CDW and SC correlations both decay as $r^{-2}$.
However, the universal correlations exponents that we find in our
exact solutions are different from these.  Furthermore, the
correlation exponents $\alpha$, $\beta$, and $\gamma$ of the FL, CDW,
and SC correlations ought to obey definite relations in a Luttinger
liquid, because they can all be written in terms of a single Luttinger
parameter $K$.  Again, the universal and non-universal correlation
exponents we find in our exact solutions do not obey these relations.
These observations bring us to the paper by Efetov and Larkin, who
first calculated the universal FL correlation exponent for an ordinary chain
of hardcore bosons to be $\alpha = \frac{1}{2}$ \cite{efetov76}.  If
we accept for the moment that the Luttinger paradigm is correct, and
that universal correlation exponents can only be found at the Fermi
liquid fixed point, then we are led to the conclusion that $\alpha =
\frac{1}{2}$ must be a correlation exponent of the Fermi liquid.
Clearly, this exponent does not belong to the Fermi liquid FL
correlation (which should be $\alpha = 1$, so what correlation does it
belong to?

In the seminal paper by Jordan and Wigner, the ordinary chain of hardcore
bosons is mapped to the ordinary chain of spinless fermions using the
Jordan-Wigner transformation (see Appendix \ref{sect:JWtrans}).  In
this transformation, the hardcore boson point operators $\bxd{j}$ and
$\bx{j+r}$ are each mapped to spinless fermion \emph{string operators}
$\cxd{j}\prod_{i<j}(-1)^{n_i}$ and $\prod_{i<j+r}(-1)^{n_i}\cx{j+r}$
respectively.  The FL correlation $\braket{\bxd{j}\bx{j+r}}$ between
two hardcore boson point operators thus become the expectation
$\braket{\cxd{j}\prod_{i=j+1}^{i=j+r-1} (-1)^{n_i} \cx{j+r}}$ of the
string operator $\cxd{j}\prod_{i=j+1}^{i=j+r-1} (-1)^{n_i} \cx{j+r}$.
which Efetov and Larkin found to decay with separation $r$ as
$r^{-1/2}$.  String correlations such as this have never been
systematically studied.  One reason for this lack of interest is that
typical string correlations, which receive contributions only
from restricted classes of configurations, decay exponentially with
$r$, as we have seen for the FL correlation in the paired limit
(Sec.~\ref{sect:tonkscorr}) and the two-leg limit
(Sec.~\ref{sect:weakXY}).  However, there appear to many string
correlations that decay with separation $r$ as power laws.  These
power law decays are associated with (quasi-)long-range order that we
have not been creative enough to imagine.  

In Sec.~\ref{sect:nointerlegGS}, we found in the two-leg limit that
the staggered ground state has long-range order, in that if we know
the $p$th particle is on leg $i = 1$, then we know for certain that
the $(p + 2s)$th particle is on leg $i = 1$, and the $(p + 2s + 1)$th
particle is on leg $i = 2$, even as $s \to \infty$, and even though we
have no idea where these particles are on the ladder.  This long-range
order is not the usual kind of long-range order, which can be written
in terms of the correlation between local order parameters, but is a
long-range \emph{string} order.  The map from the ordinary chain ground
state to the staggered ladder ground state, which is the inverse of
the one constructed in Sec.~\ref{sect:nointerlegGS}, implicitly
involves string operators, in that if we take the $p$th particle in
the ordinary ground state configuration, we will know whether to map it to
a particle on leg $i = 1$ or leg $i = 2$, \emph{after} we know which
legs the preceding particles are on.  Also, while it is deceptively
simple to describe what the string operator in this inverse map does,
which is to project out any combination of more than or equal to two
consecutive particles on the same leg of the ladder, we know of no
compact way to write down the string operator, even in this simple
limit, unlike for the case of the Jordan-Wigner string.

What we do know, drawing parallels from the Jordan-Wigner map from
hardcore bosons on ordinary chains to noninteracting spinless fermions, is
that a string map from one model to another will map some products of
local operators to string operators, for example, the hardcore boson
$\bxd{j}\bx{j+r}$ to the spinless fermion
$\cxd{j}\prod_{j'=j+1}^{j'=j+r-1}(-1)^{n_{j'}}\cx{j+r}$, and other
products of local operators to products of local operators, for
example, the hardcore boson $n_j n_{j+r}$ to the spinless fermion $n_j
n_{j+r}$.  Having understood this, we realized that the CDW$+$ and
SC$+$ correlations in the staggered ground state get mapped to the the
correlation of local operators, because the string operators involved
in the map multiply and cancel each other.  On the other hand, when we
map the CDW$-$ and SC$-$ staggered ground-state correlations to
correlations of a chain of noninteracting spinless fermions, the
string operators involved in the map do not cancel each other, and
thus the resulting ordinary chain spinless-fermion correlations are string
correlations.  We also realized that these string correlations are
operationally defined by the intervening-particle expansions we used
to compute them.

Since all the exact solutions we have obtained in this paper
can ultimately be mapped to the one-dimensional Fermi sea, we
conjecture that all correlation exponents are universal.  We claim
that: (i) all exponents that are explicitly universal are simple rational
polynomials of the single universal spinless Fermi liquid parameter $K
= 1$; and (ii) non-universal exponents are the result of (under)fitting
linear combinations of universal power laws to a single power law.
For example, in the two-leg limit, the leading universal exponent
$\beta_1 = \frac{1}{2}$ of the CDW$-$ correlation in the staggered
ground state can be shown using a bosonization calculation of the
string correlation it is mapped to, to follow automatically from the
universal Fermi liquid parameter $K = 1$ \cite{henley05}.  In this
same limiting case, the leading universal correlation exponent
$\gamma_1 = \frac{5}{2}$ of the SC$-$ correlation, which gets mapped
to a significantly more complicated string correlation, can
conceivably be written as the combination
\begin{equation}
2K + \frac{1}{2K} = \frac{5}{2} 
\end{equation} 
of the universal Fermi liquid parameter $K = 1$, even though the bosonized
form of this string correlation is not known.  For the excluded chain of
hardcore bosons or spinless fermions, nonlinear curve fitting of the
SC correlation to the sum of one leading power-law decay and one
subleading power-law decay leads to weakly non-universal correlation
exponents for both power laws, whereas a complicated sum of power-law
decays, Eq.~\eqref{eqn:SCmixed}, produces a better fit visually.  We
believe good fits can also be obtained, using similar complicated sum
of power-law decays, for those numerical correlations which we found
to have strongly non-universal correlation exponents.

Finally, we asked ourselves whether all these string correlations that
we have predicted will decay with separation $r$ slower than the
two-point function $\braket{\cxd{j}\cx{j+r}} \sim r^{-1}$ can be
measured in a chain of noninteracting spinless fermions.  Since these
string operators are nonlocal observables, they do not in general
couple to local measurements, so direct experimental measurement would
be challenging, if not downright impossible.  However, we would like
to suggest the following possibility: for a given string correlation of
the one-dimensional Fermi sea, cook up in the laboratory an
experimental system in which the corresponding correlation is a point
correlation.  If the ground-state of the experimental system can be
mapped to the one-dimensional Fermi sea, we expect a measurement of
the point correlation exponent in the experimental system to be an
indirect measurement of the string correlation exponent in the Fermi
sea. 

\begin{appendix}

\section{Jordan-Wigner Transformation}
\label{sect:JWtrans}

On a one-dimensional chain, hardcore bosons cannot move past each
other, as one boson must first hop on top of the other --- a move
explicitly forbidden by the hardcore condition --- for this to happen.
For a different reason (the Pauli Exclusion Principle), but to the
same effect, noninteracting spinless fermions on a one-dimensional
chain cannot exchange positions.  Therefore, in one dimension, the
hardcore-boson and noninteracting-spinless-fermion Hamiltonians are
also identical in structure, and thus the ground state of a chain of
hardcore bosons is related to the Fermi-sea ground state of a chain
of noninteracting spinless fermions in a simple way.  A translation
machinery exists to map back and forth between these two ground
states.  This is the Jordan-Wigner transformation \cite{jordan28}
\begin{equation}\label{eqn:jordanwigner}
\bx{i} = \prod_{j < i}(\one - 2\cxd{j}\cx{j})\,\cx{i}, \quad
\bxd{i} = \cxd{i}\prod_{j < i}(\one - 2\cxd{j}\cx{j}),
\end{equation}
which maps hardcore bosons to spinless fermions, where the product
\begin{equation}
\prod_{j < i} (\one - 2\cxd{j}\cx{j}) = \prod_{j < i} (\one - 2 n_j) =
\prod_{j < i} (-1)^{n_j}
\end{equation}
is called the \emph{Jordan-Wigner string}.

In Section \ref{sect:infinitelystrongcorrelatedhops} we saw how pairs
of spinless fermions bound by correlated hops in the limit $t' \gg
t_{\perp}, t_{\parallel}$ can be mapped to hardcore bosons with
infinite nearest-neighbor repulsion, and then to hardcore bosons using
the right-exclusion map described in Section \ref{sect:nnetonni}, and
then finally to noninteracting spinless fermions.  In Section
\ref{sect:nnegsexp}, we saw how excluded hardcore-boson expectations
are related to appropriately chosen ordinary hardcore-boson
expectations.  This relation between excluded hardcore-boson
expectations and ordinary hardcore-boson expectations will typically
involve the intervening-particle expansion
Eq.~\eqref{eqn:interpartexp}.  As such, we will encounter
hardcore-boson expectations of the form
\begin{equation}\label{eqn:hcbexp}
\braket{\bxd{i}(\one - n_{i+1})\cdots n_{i+l_1} \cdots n_{i+l_p} \cdots
(\one - n_{i+r'-1})\bx{i+r'}},
\end{equation}
a lot, where there are $p$ hardcore-boson occupation number operators
$n_{i+l}$, at sites $i+l$, and $r'-p-1$ hardcore-boson operators
$(\one - n_{i+l'})$, at sites $i+l'$, between the hardcore boson
operators $\bxd{i}$ at site $i$ and $\bx{i+r'}$ at site $i + r'$.

To evaluate these expectations, we first invoke the Jordan-Wigner
transformation \eqref{eqn:jordanwigner} to replace all the hardcore-boson 
occupation number operators $n_j = \bxd{j}\bx{j}$ by spinless-fermion 
occupation number operators $n_j = \cxd{j}\cx{j}$ in \eqref{eqn:hcbexp}.
Then, to account for the two unpaired hardcore-boson operators at the ends 
of the hardcore-boson operator product, we write \eqref{eqn:hcbexp} as the
spinless-fermion expectation
\begin{multline}
\langle\cxd{i}\prod_{j < i}(\one - 2 n_j)
(\one - n_{i+1})\cdots n_{i+l_1} \cdots n_{i+l_p} 
\cdots (\one - n_{i+r'-1}) \times {} \\
\prod_{j < i} (\one - 2 n_j) \prod_{i \leq j < i + r'} (\one - 2 n_j)\,
\cx{i+r'}\rangle.
\end{multline}
Noting that all Jordan-Wigner string operators $(\one - 2n_j)$ commutes 
with $n_{j'}$ and $(\one - n_{j'})$, for $j < i$ and $i < j' < i + r'$, 
and that
\begin{equation}
(\one - 2n_j)(\one - 2n_j) = \one,
\end{equation}
we can bring the Jordan-Wigner string $\prod_{j < i}(\one - 2 n_j)$
associated with the annihilation operator $\cx{i+r'}$ through the 
intervening spinless-fermion operators to obtain
\begin{equation}
\langle\cxd{i} (\one - n_{i+1})\cdots n_{i+l_1} \cdots n_{i+l_p} 
\cdots (\one - n_{i+r'-1})
\prod_{i \leq j < i + r'} (\one - 2 n_j)
\cx{i+r'}\rangle.
\end{equation}
Then, using the fact that
\begin{equation}
\cxd{i}(\one - 2 n_i) = \cxd{i}, \quad
n_j (\one - 2 n_j) = -n_j, \quad
(\one - n_j) (\one - 2 n_j) = (\one - n_j),
\end{equation}
we can finally write the hardcore-boson expectation
\begin{equation}\label{eqn:spinlessfermionexpectation}
\braket{\bxd{i}\prod_{\text{empty}}(\one - n_j)
\prod_{\text{filled}} n_j\, \bx{i+r'}} = (-1)^p
\braket{\cxd{i}\prod_{\text{empty}}(\one - n_j)
\prod_{\text{filled}} n_j\, \cx{i+r'}}
\end{equation}
as a spinless-fermion expectation, where $p$ is the number of occupied
sites between $i$ and $i + r'$.  The suffixes `empty' or `filled' in 
the products in \eqref{eqn:spinlessfermionexpectation} refer to the 
sites between $i$ and $i + r$ which are empty or filled respectively.

\end{appendix}

\begin{acknowledgments}
This research is supported by NSF grant DMR-0240953, and made use of
the computing facility of the Cornell Center for Materials Research
(CCMR) with support from the National Science Foundation Materials
Research Science and Engineering Centers (MRSEC) program
(DMR-0079992).  SAC also acknowledge support from the the Nanyang
Technological University startup grant SUG 19/07.
\end{acknowledgments}


\begin{thebibliography}{99}

\bibitem{tomonaga50} S. Tomonaga, Prog. Theor. Phys. \textbf{5}, 544
(1950).

\bibitem{luttinger63} J. M. Luttinger, J. Math. Phys. \textbf{4}, 1154
(1963).

\bibitem{luther74} A. Luther and V. J. Emery, Phys. Rev. Lett.
\textbf{33}, 589 (1974).

\bibitem{emery79} V. J. Emery, in \textsl{Highly Conducting
One-Dimensional Solids} edited by J. T. Devreese and R. P. Evrard and
V. E. van~Doren, pp. 247--303, Plenum Press (New York), 1979.

\bibitem{cheong07} S.-A. Cheong, Ph.D. dissertation, Cornell
University, 2007: \url{http://hdl.handle.net/1813/11559}.

\bibitem{zhang01} C. L. Henley and N. G. Zhang, Phys. Rev. B
\textbf{63}, 233107 (2001).

\bibitem{cheong09} S.-A. Cheong and C. L. Henley, Phys. Rev. B
\textbf{79}, 212402 (2009).

\bibitem{muender09} W. M\"under, A. Weichselbaum, J. von Delft, and C.
L. Henley, unpublished.

\bibitem{fendley03} P. Fendley, B. Nienhuis, and K. Schoutens, J.
Phys. A: Math. Gen. \textbf{36}, 12399 (2003).

\bibitem{efetov76} K. B. Efetov and A. I. Larkin, Sov. Phys. JETP
\textbf{42}, 390 (1976).

\bibitem{luttinger60} J. M. Luttinger, Phys. Rev. \textbf{119}, 1153
(1960).

\bibitem{oshikawa00} M. Oshikawa, Phys. Rev. Lett. \textbf{84}, 3370
(2000).

\bibitem{dzyaloshinskii03} I. Dzyaloshinskii, Phys. Rev. B
\textbf{68}, 085113 (2003).

\bibitem{korshunov03} M. M. Korshunov and S. G. Ovchinnikov, Phys.
Sol. State \textbf{45}, 1415 (2003).

\bibitem{paramekanti04} A. Paramekanti and A. Vishwanath, Phys. Rev. B
\textbf{70}, 245118 (2004).

\bibitem{liu05} Y.-L. Liu, Phys. Rev. B \textbf{72}, 155104 (2005).

\bibitem{henley05} C. L. Henley, unpublished.

\bibitem{cheong04b} S.-A. Cheong and C. L. Henley, Phys. Rev. B
\textbf{69}, 075112 (2004).

\bibitem{solyom79} J. S\'olyom, Adv. Phys. \textbf{28}, 201 (1979).

\bibitem{voit92} J. Voit, Phys. Rev. B \textbf{45}, 4027 (1992).

\bibitem{voit95} J. Voit, Rep. Prog. Phys. \textbf{58}, 977 (1995).

\bibitem{luther75} A. Luther and I. Peschel, Phys. Rev. B \textbf{12},
3908 (1975).

\bibitem{jordan28} P. Jordan and E. Wigner, Z. Phys. \textbf{47}, 631
(1928).

\bibitem{dennijs89} M. P. M. den Nijs and K. Rommelse, Phys. Rev. B
\textbf{40}, 4709 (1989).

\end{thebibliography}
\end{document}